\documentclass{aa}

\usepackage{natbib}
\bibpunct{(}{)}{;}{a}{}{,} 
\usepackage{graphicx}   
\usepackage{txfonts}
\usepackage[colorlinks=true,allcolors=cyan]{hyperref}

\begin{document}

\titlerunning{Clusters x Galaxies X Color}
\title{Cross-correlations between X-ray clusters and the general galaxy population}
\authorrunning{Comparat et al.}
\author{
Johan Comparat\inst{1,2}\thanks{E-mail: comparat@mpe.mpg.de}, 
Juan Macias-Perez\inst{2}, 
C\'eline Combet\inst{2}, 
Andrea Merloni\inst{1}, 
Dominique Eckert\inst{3},
Mohammadreza Ayromlou\inst{4},
Kirpal Nandra\inst{1},
Riccardo Seppi\inst{3}.
}

\institute{
Max-Planck-Institut f\"{u}r extraterrestrische Physik (MPE), Gie{\ss}enbachstra{\ss}e 1, D-85748 Garching bei M\"unchen, Germany
\and
Univ. Grenoble Alpes, CNRS, Grenoble INP, LPSC-IN2P3, 53, Avenue des Martyrs, 38000, Grenoble, France
\and
Department of Astronomy, University of Geneva, Ch. d'Ecogia 16, CH-1290 Versoix, Switzerland
\and
Argelander-Institut f\"ur Astronomie, Auf dem H\"ugel 71, D-53121 Bonn, Germany
}
    
\date{\today}

\abstract{
This study presents highly precise measurements of the cross-correlation between volume-limited galaxy samples from the DESI legacy survey catalogue and X-ray selected galaxy clusters from eROSITA, allowing for detailed analysis across redshift and color. Two key findings emerge. First, the cluster-galaxy cross-correlation, when split into quiescent and star-forming galaxies, contains significant information about the infall, feedback, and quenching processes of blue cloud galaxies in massive environments. These results align well with existing galaxy evolution models for higher stellar masses ($\log_{10}(M^*[M_\odot]) > 10.75$), though the red fraction may be slightly underestimated in the intermediate mass range ($10.25 < \log_{10}(M^*[M_\odot])< 10.75$). Second, the integral of the cross-correlation within 500 kpc enables a model-independent measurement of the red sequence and its scatter in clusters, providing a robust alternative to existing red-sequence calibration methods without requiring spectroscopic redshifts or classifications of galaxies. Similar analyses on upcoming photometric surveys like Euclid and LSST and spectroscopic samples like 4MOST and DESI should lead to a significant increase in the signal-to-noise ratio, particularly at small separations. }

\maketitle

\section{Introduction} 

The galaxy red sequence is a tight correlation observed in color-magnitude diagrams of galaxies, where more luminous galaxies tend to be redder. It represents a population of massive, old, and quiescent (non-star-forming) galaxies, primarily composed of older stars that have exhausted most of their gas for star formation \citep{BellWolfMeisenheimer_2004ApJ...608..752B, NelanSmithHudson_2005ApJ...632..137N, FaberWillmerWolf_2007ApJ...665..265F}. 
Red sequence galaxies dominate the population in clusters of galaxies, particularly in the densest central regions. 
In particular, detecting and measuring photometric redshift of galaxy clusters using the red sequence technique is efficient and widely used \citep{GladdersYee_2000AJ....120.2148G,KoesterMcKayAnnis_2007ApJ...660..221K,RykoffRozoBusha_2014ApJ...785..104R,ClercKirkpatrickFinoguenov_2020MNRAS.497.3976C,KlugeComparatLiu_2024A&A...688A.210K}. 

The definition of the red sequence in galaxy clusters stems from the combination of photometric and spectroscopic surveys. With the photometric survey we evaluate the color of each galaxy, while with the spectroscopic survey we measure its redshift and decide if it belongs to a cluster. 
The average color of galaxies in clusters is red and only shows a small dispersion. Importantly, the value of the color directly correlates with redshift with a small scatter, so the measurement of the color can be used as a proxy for redshift.
Central to the definition of red sequence models (and their scatter) is their need for (i) stellar population synthesis models, (ii) highly complete spectroscopic follow-up, (iii) iteration to evaluate membership that includes clipping of probable members that are outliers in color \citep{RykoffRozoBusha_2014ApJ...785..104R}. 
Such training sets are not necessarily available at all redshifts. 

In this article, we present a method to retrieve the red sequence as a function of redshift via cross-correlation between clusters and galaxies. 

The cross-correlation between galaxies and clusters has been widely used. 
Early on, \citet{Peebles_1974ApJS...28...37P} and \citet{LiljeEfstathiou_1988MNRAS.231..635L} measured a significant angular cross-correlation between Abell clusters and galaxy catalogs. 
Later, using the APM survey of galaxies and galaxy clusters, 
\citet{MaddoxSutherlandEfstathiou_1990MNRAS.243..692M}, \citet{DaltonMaddoxSutherland_1997MNRAS.289..263D}, \citet{CroftDaltonEfstathiou_1999MNRAS.305..547C} inferred the linear bias of the galaxy population and the matter density. 
Then, \citet{SanchezLambasBohringer_2005MNRAS.362.1225S} measured the cross-correlation between X-ray selected clusters \citep[REFLEX,][]{BohringerSchueckerGuzzo_2004A&A...425..367B} and galaxies from APM and 2MASS \citep{JarrettChesterCutri_2000AJ....119.2498J}, the earliest precursor to this study. They measured with significance the presence of both the one-halo and two-halo terms in the cross-correlation signal and how it increases with the X-ray luminosity threshold.
This measurement (and more generally any cross-correlation summary statistic, \citealt{BanerjeeAbel_2021MNRAS.504.2911B}) was shown to be sensitive to cosmological parameters \citep{HutsiLahav_2008A&A...492..355H, FedeliCarboneMoscardini_2011MNRAS.414.1545F,SalcedoWibkingWeinberg_2020MNRAS.491.3061S} and of interest for upcoming surveys such as Euclid and LSST \citep{LaureijsAmiauxArduini_2011arXiv1110.3193L,EuclidCollaborationMellierAbdurro'uf_2025A&A...697A...1E, IvezicKahnTyson_2019ApJ...873..111I}. 
\citet{ZuWeinberg_2013MNRAS.431.3319Z} measured and modeled the cross-correlation between optically selected galaxy groups \citep{YangMovandenBosch_2007ApJ...671..153Y} and SDSS galaxies \citep{YorkAdelmanAnderson_2000AJ....120.1579Y}. 
They constrain their model of galaxy infall kinematics (GIK), arguing that the large-scale infall of galaxies around massive clusters may be a powerful diagnostic of structure growth, dark energy, or deviations from general relativity. 
\citet{RobertsonHuffMarkovic_2024MNRAS.533.4081R} recently refined the GIK model using the INDRA simulations \citep{FalckWangJenkins_2021MNRAS.506.2659F} for possible precision cosmology inference. 
\citet{PaechHamausHoyle_2017MNRAS.470.2566P} inferred the large-scale halo bias of optically-selected galaxy clusters \cite{RykoffRozoBusha_2014ApJ...785..104R} by cross-correlating them (in Fourier space) with photometric galaxies from SDSS \citep{YorkAdelmanAnderson_2000AJ....120.1579Y}. 

{Although the signal-to-noise ratio in the summary statistics previously measured was high, only a few studies explored a possible trend as a function of galaxy properties, like colors or star formation rate \citep[e.g.][]{ BaxterChangJain_2017ApJ...841...18B, AdhikariShinJain_2021ApJ...923...37A}. 
In addition, recent lensing surveys measured the profile of red-sequence and blue cloud galaxies in clusters \citep{HennigMohrZenteno_2017MNRAS.467.4015H, Nishizawa2018PASJ...70S..24N}. }
In a sense, these profile measurements are a cross-correlation between cluster and galaxy positions. 
They measure the dominance of red galaxies in the inner parts of clusters. 

This article further explores this parameter space with the cross-correlation measurement as a tool. 
This attempts to go beyond the analysis of galaxy fractions as a function of cluster-centric radius \citep[e.g.][]{vanderWel_2008ApJ...675L..13V,BianconiSmithHaines_2020MNRAS.492.4599B, VulcaniPoggiantiGullieuszik_2023ApJ...949...73V}.

In addition, a related cross-correlation between a hot gas measure (X-ray or SZ) and galaxies is a well-known signpost of feedback \citep{DaviesCrainMcCarthy_2019MNRAS.485.3783D,KukstasMcCarthyBaldry_2020MNRAS.496.2241K,TruongPillepichWerner_2020MNRAS.494..549T,TruongPillepichWerner_2021MNRAS.501.2210T,MoserBattagliaNagai_2022ApJ...933..133M,GraysonScannapiecoDave_2023ApJ...957...17G,LauNagaiBogdan_2024arXiv241204559L, SoriniBoseDave_2024OJAp....7E.115S}. 
Such cross-correlations measurements, aimed to constrain galaxy evolution models using the X-ray or SZ field, get refined with progressing observations in SZ \citep{YanvanWaerbekeTroster_2021A&A...651A..76Y,AmodeoBattagliaSchaan_2021PhRvD.103f3514A,IbitoyeTramonteMa_2022ApJ...935...18I, PandeyGattiBaxter_2022PhRvD.105l3526P, SanchezOmoriChang_2023MNRAS.522.3163S,DasChiangMathur_2023ApJ...951..125D,OrenSternbergMcKee_2024ApJ...974..291O} and X-rays \citep{ComparatTruongMerloni_2022A&A...666A.156C,ZhangComparatPonti_2024A&A...690A.267Z, ZhangComparatPonti_2024A&A...690A.268Z, ZhangComparatPonti_2025A&A...693A.197Z,LaPostaAlonsoChisari_2024arXiv241212081L,LiFangGe_2024ApJ...977L..40L, PopessoMariniDolag_2024arXiv241117120P,DevDriverMeyer_2024MNRAS.535.2357D,ComparatMerloniPonti_2025A&A...697A.173C}.

We focus here on the cross-correlation between X-ray-selected galaxy clusters and galaxies. 
We use the eROSITA cluster catalogs derived from its first all-sky survey \citep[eRASS1,][]{MerloniLamerLiu_2024A&A...682A..34M,BulbulLiuKluge_2024A&A...685A.106B}. 
eROSITA (extended ROentgen Survey with an Imaging Telescope Array) is a wide-field X-ray telescope on board the Russian-German "Spectrum-Roentgen-Gamma" (SRG) observatory, currently the best X-ray instrument to construct cluster samples \citep{MerloniPredehlBecker_2012arXiv1209.3114M,PredehlAndritschkeArefiev_2021A&A...647A...1P,SunyaevArefievBabyshkin_2021A&A...656A.132S}. 
We combine it with galaxies selected from the legacy survey  \citep{DeySchlegelLang_2019AJ....157..168D}, which is currently the best match in depth and area covered to the eROSITA cluster sample. 

In this article, we push further the statistical study of the cross-correlation between clusters and galaxies as a function of galaxy color. 
We describe the observations used in Section \ref{sec:data} and the cross-correlation measurements in Section \ref{sec:measurements}. We discuss our findings in Section \ref{sec:discussion}.

\section{Observations}
\label{sec:data}

In this section, we describe the X-ray-selected cluster catalogs (Sec. \ref{subsec:data:xray}) and the galaxy catalogs (Sec. \ref{subsec:data:galaxies}).

\begin{table*}
\begin{center}
\caption{{Description of the galaxy and cluster sample combinations considered.} }
\label{tab:Vlim:samples:basic:info}
\begin{tabular}{l rrrrrrrrrrrrrrrr}
\hline \hline
name & \multicolumn{4}{c}{Cluster samples} & \multicolumn{4}{c}{Galaxy samples} & Section \\
& z min & z max &  $\log_{10}\left(\frac{L_X}{[erg/s]}\right)$ & N & z min & z max & $\log_{10}\left(\frac{M^*}{[M_\odot]}\right)$ & N  \\
\hline

C0 x G1025 & 0.1 & 0.2 & 42.7 & 1,650 & 0.1 & 0.2 & 10.25 &  2,255,070 & \ref{subsec:measurements:rsbc} \\
\hspace*{0.2cm} \& RS &  &   &  & & & &  &  1,438,690 & \\
\hspace*{0.2cm} \& BC &  &   &  & & & &  &    538,145 & \\
C1 x G1075 & 0.1 & 0.3 & 43.1 & 2,993 & 0.1 & 0.3 & 10.75 &  2,480,941 & \ref{subsec:measurements:rsbc} \\
\hspace*{0.2cm} \& RS &  &   &  & & & &  &  1,871,758 & \\
\hspace*{0.2cm} \& BC &  &   &  & & & &  &    383,972 & \\
C2 x BGS   & 0.1 & 0.4 & 43.4 & 3,732 & -    &  -   & -     & 13,881,761 &  \ref{subsec:measurements:colors} \\
\hline
\end{tabular}
\tablefoot{
{N gives the total number of sources present in each sample. The quantities in the  $\log_{10}\left(\frac{L_X}{[erg/s]}\right)$ and $\log_{10}\left(\frac{M^*}{[M_\odot]}\right)$ columns correspond to the minimum selection threshold applied.}
}
 \end{center}
\end{table*}

\subsection{eROSITA, eRASS:1 volume limited cluster sample}
\label{subsec:data:xray}

eROSITA, with its seven identical Wolter-1 mirror modules, is a sensitive wide-field X-ray telescope capable of delivering deep, sharp images over vast sky areas in the energy range of 0.2-8 keV, with maximum sensitivity in the 0.3-2.3 keV range. 
In its first survey, released in 2024\footnote{\url{https://erosita.mpe.mpg.de/dr1}} \citep{MerloniLamerLiu_2024A&A...682A..34M}, the German eROSITA consortium observed the Western Galactic hemisphere. 
About 12,247 clusters were detected as extended sources in X-rays and optically confirmed \citep{BulbulLiuKluge_2024A&A...685A.106B}. Their photometric redshifts were measured using redmapper in scan mode \citep{RykoffRozoBusha_2014ApJ...785..104R,KlugeComparatLiu_2024A&A...688A.210K}. 
\citet{SeppiComparatGhirardini_2024A&A...686A.196S} extracted volume-limited samples from this flux-limited survey to perform a detailed clustering analysis. 
We use three of these volume-limited cluster (and random) catalogs: the C0, C1, and C2 samples (containing each 1650, 2993, 3732 sources) that cover redshifts above 0.1 and below 0.2 (0.3, 0.4) and have $\log_{10}$ soft X-ray luminosities larger than 42.7, 43.1, 43.4. 
The mean redshifts of the samples are 0.153, 0.206, 0.264. 
The mean X-ray luminosity is $\bar{L}_X$=4.63$\pm$0.97 (7.38$\pm$0.98, 10.21$\pm$0.98) $\times 10^{43}$ erg s$^{-1}$. 
The purity of the samples is larger than 95\%. 
\citet{SeppiComparatGhirardini_2024A&A...686A.196S} measured and modeled (with halo occupation distribution, \citealt{CooraySheth_2002PhR...372....1C, MoreMiyatakeMandelbaum_2015ApJ...806....2M}) the correlation functions of these samples. 
They find satellite cluster fractions consistent with zero: $f_{sat}$ <14.9\% (<9.3\%, <9.9\%), showing that the cluster samples are mainly constituted of distinct halos. 
The deduced large-scale halo bias increases with the luminosity threshold (and mean luminosity) applied $b=2.95\pm$0.21 (3.35$\pm$0.23, 3.69$\pm$0.27), see their Fig. 8. Said differently, the mean halo mass, $\bar{M}_{200m}$=3.09$\pm$0.48, (3.54$\pm$0.51, 3.83$\pm$0.55) $\times 10^{14}$ M$_\odot$, correlates with the average X-ray luminosity. 
These characteristics set the scene for the clusters fed into the cross-correlation.

The choice of the cluster samples depends on the available galaxy samples. As detailed below, the galaxy samples considered here extend to a maximum redshift of 0.4. This limit justifies our focus on these three low-redshift cluster samples, not higher redshift samples. 
 
\subsection{Legacy survey DR10.1 galaxies}
\label{subsec:data:galaxies}

We use the legacy survey tenth data release (LS10 hereafter) to select a flux-limited galaxy sample \citep{DeySchlegelLang_2019AJ....157..168D}. This sample is described in detail in \citet{ComparatMerloniPonti_2025A&A...697A.173C}. 
We summarize here the main information. LS10 provides source catalogs for Dec$<$32$^\circ$, from which we select low redshift galaxies using an algorithm similar to that of the DESI BGS \citep{HahnWilsonRuiz-Macias_2023AJ....165..253H} with a magnitude cut $r_{AB}<19.5$. 
The main difference between our re-implementation of the BGS selection and the original one is that we are stricter on the MASKBITS and color cuts to increase the purity of the photometric sample and guarantee an unbiased clustering estimate when using photometric redshifts. 
Using the BGS spectroscopic sample will naturally remove contaminants so their cuts can be slightly looser, increasing their completeness without impacting their purity. 
Unfortunately, in the Western Galactic hemisphere observed by the German eROSITA consortium, only limited spectroscopy is available, so we need to be stricter in the definition of the galaxy sample. 
Following the BGS selection criteria, we separate stars from galaxies and select galaxies with $13<r_{AB}<19.5$ to obtain a set of BGS-like galaxies.
We require the galactic reddening to be small (E(B-V)$<0.1$) and to have at least one observation in each of the $g,r,z$ bands. 
With mask bits, we remove secondary detections (\texttt{MASKBIT} 0), sources that touch Tycho sources with \texttt{MAG\_VT} $<$ 13, and Gaia stars with G $<$ 13 (\texttt{MASKBIT} 1), and sources touching a pixel in a globular cluster (\texttt{MASKBIT} 13).
We also remove sources with the \texttt{FITBITS} 1, 6, 10, 11, 12, and 13 turned on.
Finally, we remove Gaia duplicates (\texttt{TYPE}="DUP"). We follow the BGS selection to discard artifacts, but with a stricter criterion in color-color space: we keep sources with $-0.2<g-r<2$ and $-0.2<r-z<1.6$.
We obtain a galaxy sample with a completeness larger than 90\%. 
The sample contains 13,881,761 sources over a footprint of 16,796 square degrees and extends up to redshift 0.5. 

\subsection{Sample combinations}

In the following, we consider two cases for the cross-correlation: \textit{(i)} clusters crossed with the general population of galaxies split in red-sequence and blue cloud, \textit{(ii)} clusters crossed with galaxies in narrow bins of colors. 
{Table \ref{tab:Vlim:samples:basic:info} summarizes the sample combinations considered.}

For the first case, to have the cleanest interpretation, we need a volume-limited set of galaxies, and we use these designed and studied in \citet{ComparatMerloniPonti_2025A&A...697A.173C}. 
We consider cluster-galaxy sample-pairs so that the galaxy sample completely covers the redshift interval of clusters. Among the possible galaxy samples, we use the one with the most significant number of galaxies, i.e., the lowest stellar mass threshold. 
Among the cluster samples, only the C0 ($0.1<z<0.2$) and C1 ($0.1<z<0.3$) cluster samples have their redshift intervals covered by the galaxy samples considered. Among them, the ones covering at least the cluster redshift range and that maximize the number of galaxies are: the sample selected with $10.25<\log_{10}(M^*[M_\odot])<12$, 0.05$<z<$0.22 (named G1025) for C0; and the one with $10.75<\log_{10}(M^*[M_\odot])<12$, 0.05$<z<$0.31 (named G1075) for C1. 
We name these combinations C0xG1025 and C1xG1075. 
In \citet{ComparatMerloniPonti_2025A&A...697A.173C}, we measured the autocorrelation function of these galaxy samples and fitted a HOD model upon it. 
The G1025 (G1075) sample is made of 3,280,777 (2,768,066) galaxies; it has a mean redshift of 0.162 (0.226) and a mean stellar mass ($\log_{10}(M^*/M_\odot)$) of 10.66 (11.02). 
{The redshift selection, $0.1<z<0.2$ ($0.1<z<0.3$), to match that of the cluster samples C0 (C1) reduces the total number of galaxies to 2,255,070 (1,480,941) for the G1025 (G1075) sample.} 
From the HOD fits, we derived a large-scale halo bias value of 1.29 (1.43) and a mean halo mass hosting central galaxies of $\log_{10}(M_{200m}/M_\odot)=12.8$ (13).

For the second case, we need the largest number of galaxies to enable the finest binning of the galaxies as a function of colors, so we use the complete sample described above (named BGS). We need a cluster sample that covers the largest redshift range in overlap with the galaxies, leading us to use sample C2 ($0.1<z<0.4$). We call this combination C2xBGS.

\section{The cross-correlation between individual galaxies and X-ray selected clusters}
\label{sec:measurements}

We describe the cross-correlation between clusters and the general population of galaxies, and when split into red-sequence and blue cloud in Sec. \ref{subsec:measurements:rsbc}. 
Then, in Sec. \ref{subsec:measurements:colors}, we investigate how the cross-correlation measurement varies with the $g-r$ color of galaxies and the cluster redshifts.

\subsection{The cross-correlation between red sequence, blue cloud galaxies, and X-ray selected clusters}
\label{subsec:measurements:rsbc}

\begin{figure*}
    \centering
    \includegraphics[width=0.9\columnwidth]{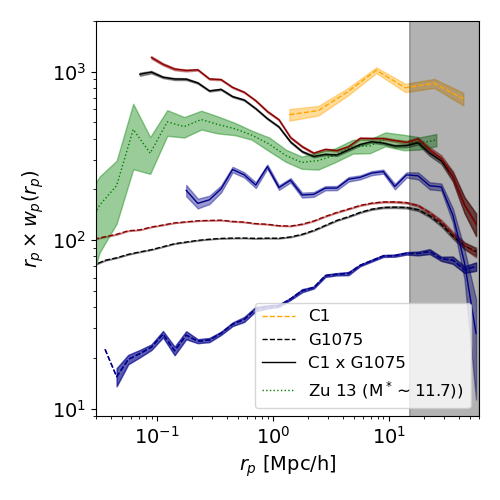}
    \includegraphics[width=0.9\columnwidth]{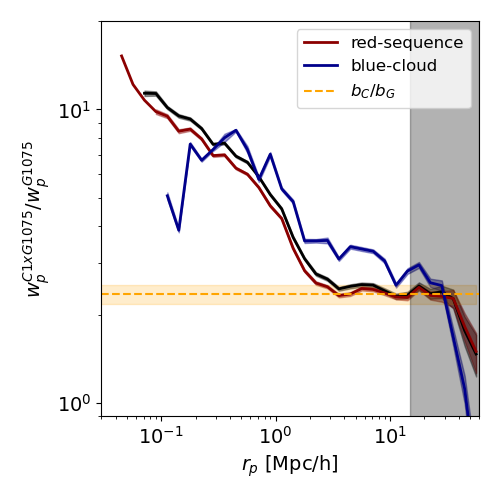}
    \includegraphics[width=0.7\columnwidth]{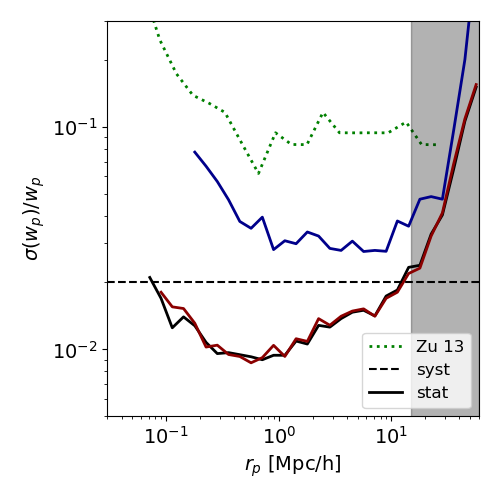}
    \includegraphics[width=0.95\columnwidth]{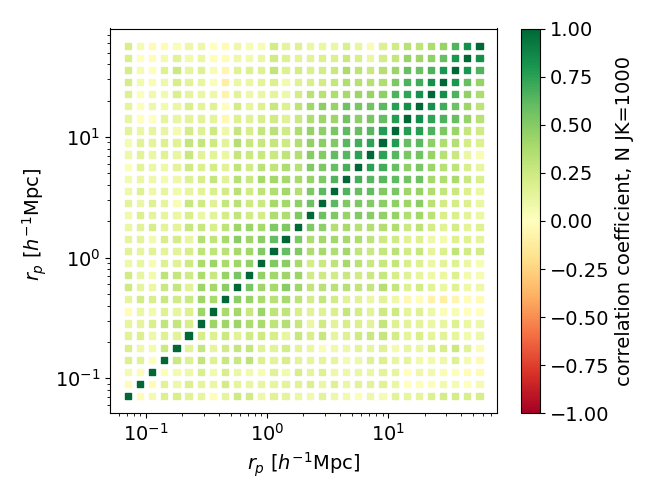}
    \caption{C1xG1075. Correlation functions measured using the C1 cluster sample ($0.1<z<0.3$, $\log_{10}(L_X\; [\mathrm{erg/s}])>43.1$) in combination with the G1075 galaxy sample ($10.75<\log_{10}(M^*[M_\odot])<12$). 
    {Left panel:}
    The projected autocorrelation of galaxies (and clusters) split into red sequence and blue cloud is shown with the dashed lines. 
    The projected cluster-galaxy cross-correlation functions are in solid lines. 
    This signal is dominated by the red sequence galaxies: the red lines are significantly higher than the blue lines. 
    The measurement from \citet{ZuWeinberg_2013MNRAS.431.3319Z} crossing the groups from \citet{YangMovandenBosch_2007ApJ...671..153Y} with a brightest group galaxy with stellar mass in the range ($11.4<\log_{10}(M^*_{BGG}[M_\odot])<11.9$) and all SDSS galaxies is shown with a green dotted line. 
    {In this analysis, we trust the measurements up to 15 $h^{-1}$Mpc, leftward of the grayed area, beyond which, it may be affected by systematics uncertainties.}
    {Right panel:}
    The ratio between the galaxy-cluster cross-correlation and the respective galaxy auto-correlations. This panel compares the strength of the clustering in the cluster environment compared to that of all galaxies (in any environment). 
    The ratio is above one, meaning galaxies cluster more near galaxy clusters. 
    For red-sequence galaxies, the extra power in clustering increases when scales decrease. 
    For blue cloud galaxies, the clustering increases towards small scales until 0.3 Mpc/h, where it stalls and turns over below 0.2 Mpc/h. 
    The green horizontal line shows the ratio of large-scale halo biases determined in \citet{SeppiComparatGhirardini_2024A&A...686A.196S} and \citet{ComparatMerloniPonti_2025A&A...697A.173C}. Following expectations, it aligns with the ratio of correlation functions on large scales. 
    {Bottom panels:}
    {Results of the Jackknife procedure. Relative uncertainties on the cross-correlation functions (left panel) and cross-correlation coefficient for the C1xG1075 cross-correlation (right panel). The measurements have relative uncertainties ranging from 1\% to 10\%. The previous measurement done with SDSS had 10\%-30\% uncertainties}. 
    Thanks to the high density of galaxies available with LS10, we gain precision in this measurement.
    }
    \label{fig:wprp:C1xG1075}
\end{figure*}

\begin{figure*}
    \centering
    \includegraphics[width=0.9\columnwidth]{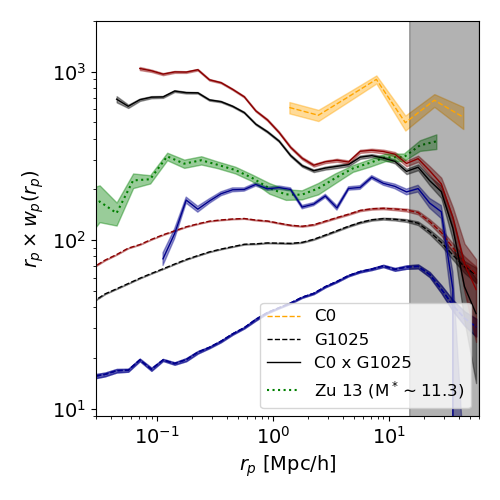}
    \includegraphics[width=0.9\columnwidth]{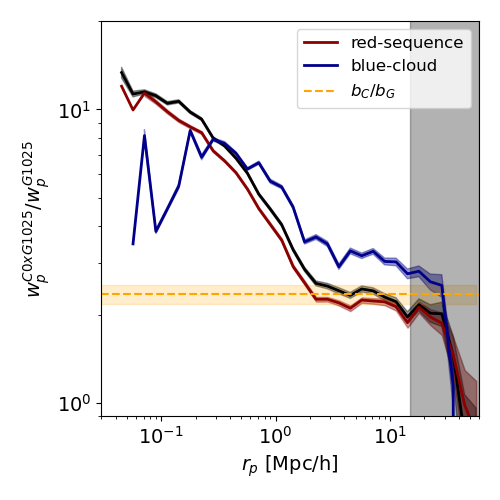}
    \includegraphics[width=0.7\columnwidth]{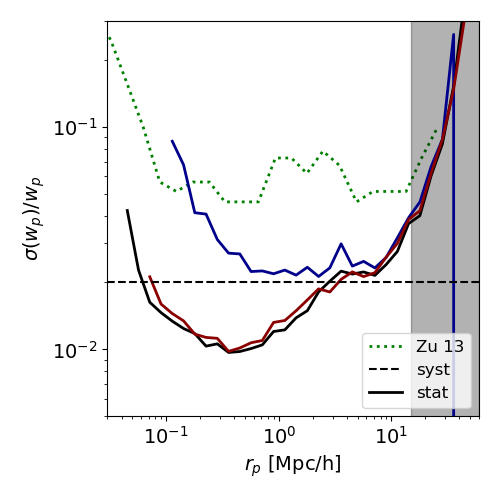}
    \includegraphics[width=0.95\columnwidth]{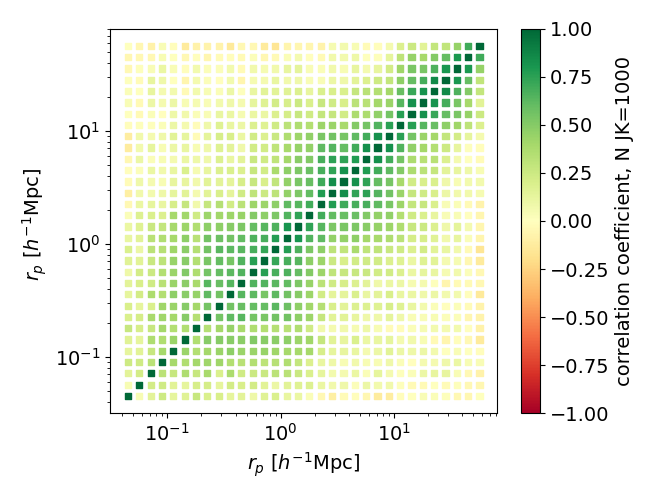}
    \caption{C0xG1025. Continued Fig. \ref{fig:wprp:C1xG1075}. Correlation functions measured using the C0 cluster sample ($0.1<z<0.2$, $\log_{10}(L_X\; [\mathrm{erg/s}])>42.7$) in combination with the G1025 galaxy sample ($10.25<\log_{10}(M^*[M_\odot])<12$)
    The measurement from \citet{ZuWeinberg_2013MNRAS.431.3319Z} crossing the groups from \citet{YangMovandenBosch_2007ApJ...671..153Y} with a brightest group galaxy with stellar mass in the range ($11.2<\log_{10}(M^*_{BGG}[M_\odot])<11.4$) and all SDSS galaxies is shown with dotted line. 
    }
    \label{fig:wprp:C0xG1025}
\end{figure*}

The projected cross-correlation, $w_p(r_p)$ is obtained with the estimator from \citet{LandySzalay_1993ApJ...412...64L} using a galaxy sample and its random sample \citep[both described in ][]{ComparatMerloniPonti_2025A&A...697A.173C} plus a cluster sample and its random sample \citep[both described in ][]{SeppiComparatGhirardini_2024A&A...686A.196S}, so two distinct data samples and two distinct random samples with each their photometric redshift measurements. 
We estimate uncertainties by repeating the measurement {1000 times, each time removing 10\% of the area (discarding a set of healpix pixels each time). 
Another difficulty with this photo-z defined galaxy sample is the control of systematic variations of the galaxy number density due to external variables. 
We find a remaining 1-2 per cent variation not captured in the random catalogues for these two samples. Addressing quantitatively these small discrepancies requires extensive spectroscopy. 
Until more spectroscopic data is available in the south (e.g., in 5 years with 4MOST \citealt{deJongAgertzBerbel_2019Msngr.175....3D}), we are left with adding a systematic uncertainty floor of 2\% on the correlation function measurements}.    

We show in Fig. \ref{fig:wprp:C1xG1075} and Fig.~\ref{fig:wprp:C0xG1025} the auto and cross-correlation functions for C1xG1075 and C0xG1025, respectively. 
The autocorrelation of galaxies split into red sequence and blue cloud is accurately measured, with sub-percent statistical uncertainties (see middle panels, dashed lines). 
The cluster-galaxy cross-correlation functions also show small relative uncertainties in the range of 2-10\% (see middle panels, solid lines). 
This is a clear improvement compared to the previous estimate from \citet{ZuWeinberg_2013MNRAS.431.3319Z}, which crossed the groups from \citet{YangMovandenBosch_2007ApJ...671..153Y} with all SDSS galaxies to obtain relative uncertainties in the range 10\%-50\%. 
The gain in signal-to-noise is due to the 5-10 times denser galaxy samples used.

When considering all galaxies, the split in red and blue is clear. In the cross-correlation, the blue galaxies depart from the average cross-correlation. Still, the red ones are almost on top of the average, meaning that the red galaxies dominate the cross-correlation signal. 
This description applies to the two sample combinations studied. We do not see a significant trend by increasing the redshift range from 0.1$<z<$0.2 to 0.1$<z<$0.3 and the luminosity threshold from 42.7 to 43.1.
These considerations are in qualitative agreement with previous studies of the occupation of red and blue galaxies in clusters \citep{HennigMohrZenteno_2017MNRAS.467.4015H, Nishizawa2018PASJ...70S..24N} and the fact that red sequence galaxies dominate in fraction at high halo masses at low redshift \citep[\textit{e.g.}][]{ZuMandelbaum_2015MNRAS.454.1161Z, ZuMandelbaum_2016MNRAS.457.4360Z}. 

Since the X-ray luminosity correlates (with 0.3 dex scatter) with halo mass, this cross-correlation directly probes the galaxy population in high-mass dark matter haloes. 
The ratio of the cross-correlation between galaxies and clusters ($w^{CxG}_p$) to the galaxy-galaxy auto-correlation ($w^{GxG}$) (see right panels of
Fig. \ref{fig:wprp:C1xG1075} and \ref{fig:wprp:C0xG1025}) inform on the extra power from the massive haloes. 
It directly probes the influence of the environment on the different classes of galaxies.

On large scales ($r_p>3\; h^{-1} \mathrm{Mpc}$), this correlation function ratio should equate the ratio of large-scale halo bias 
\begin{equation}
\frac{w^{CxG}_p}{w^{GxG}_p}\; (r_p>3\; h^{-1} \mathrm{Mpc})\; \propto \frac{b_{G}b_{C}}{b^2_{G}}=\frac{b_{C}}{b_{G}}.
\end{equation}
Indeed, we find that the correlation function ratio is in agreement with the ratio of large-scale halo biases inferred by \citep{SeppiComparatGhirardini_2024A&A...686A.196S} for the clusters and by \citet{ComparatMerloniPonti_2025A&A...697A.173C} for the galaxies (see the green horizontal shaded area in the right panels of Figs. \ref{fig:wprp:C1xG1075} and  \ref{fig:wprp:C0xG1025}). 

Let's focus now on the smaller scales inside the 1-halo term. 
For the red sequence galaxies, the ratio between cross-correlation and auto-correlation steadily increases when separation decreases, from 2$h^{-1}$Mpc down to 50$h^{-1}$kpc. 
For blue cloud galaxies, the ratio between cross-correlation and auto-correlation increases towards smaller separation from 2$h^{-1}$Mpc until 0.3$h^{-1}$Mpc, where it stalls and turns over below 0.2$h^{-1}$Mpc. 
This turnover corresponds to the absence of blue cloud galaxies towards the cluster centers. Thus, we find evidence for an infalling population of blue-cloud galaxies that eventually get ram-pressure stripped and disappear when they enter the clusters. 

\subsubsection*{Comparison with models}

\begin{figure*}
    \centering
    \includegraphics[width=0.47\linewidth]{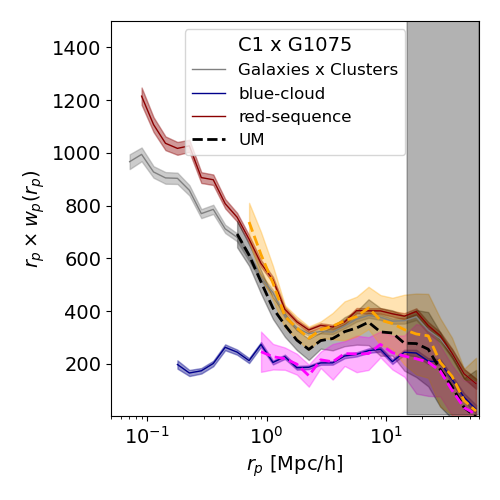}
    \includegraphics[width=0.47\linewidth]{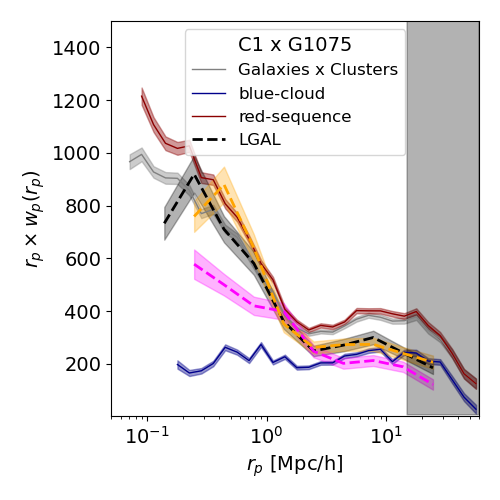}
    \includegraphics[width=0.47\linewidth]{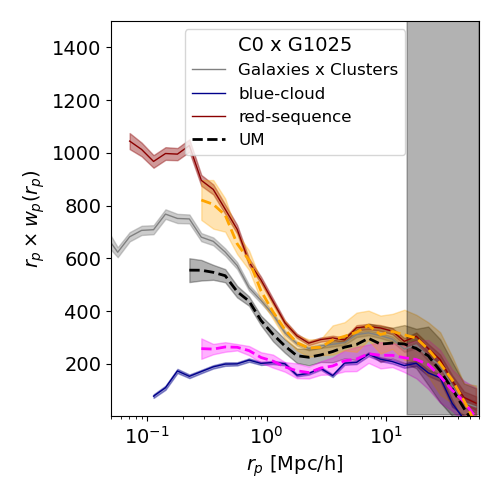}
    \includegraphics[width=0.47\linewidth]{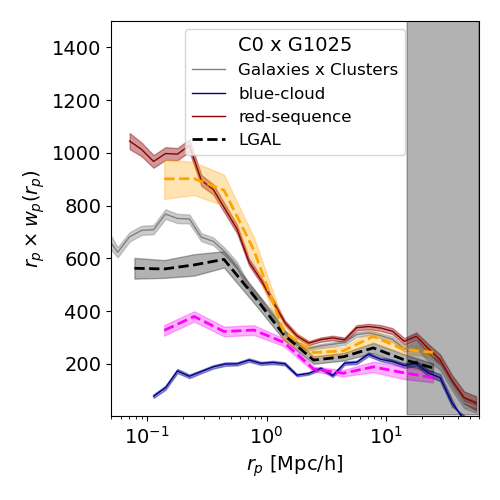}
    \caption{We show the Uchuu+UniverseMachine (UM) model \citep{AungNagaiKlypin_2023MNRAS.519.1648A} (left column of panels) and the LGAL model \citep{AyromlouKauffmannYates_2021MNRAS.505..492A} (right column of panels) compared to the C1xG1075 (C0xG1025) galaxy-cluster cross-correlation in the top (bottom) panel. 
    The model prediction for red-sequence (blue-cloud) galaxies is shown in orange (pink). 
    The UM model agrees well with C1xG1075 observations. 
    At lower redshift, with lower stellar mass galaxies, the UM model shows a hint of discrepancy with the observations, hinting that the quiescent fraction of galaxies with stellar mass of $\sim$10.5 in clusters may be underestimated. 
    The LGAL model agrees with the measurements of the complete galaxy population (grey lines) and red galaxies (red and orange lines) for the C1xG1075 and the C0xG1025 down to 0.5 Mpc/h. 
    The LGAL blue-cloud galaxies show excess power on small scales compared to the observations.}
    \label{fig:CxG:UM}
\end{figure*}

We compare these results with a prediction constructed with the publicly available \textsc{UniverseMachine} model from \citep{BehrooziWechslerHearin_2019MNRAS.488.3143B,AungNagaiKlypin_2023MNRAS.519.1648A} run on the Uchuu simulation \citep{IshiyamaPradaKlypin_2021MNRAS.506.4210I}. 
We use the two snapshots near the mean redshifts of the samples considered ($z=$0.14, $z=$0.19), apply the same cuts as in observations, and predict the equivalent summary statistic, see Fig. \ref{fig:CxG:UM}. 
To do so, we apply the method of \citet{ComparatEckertFinoguenov_2020OJAp....3E..13C} and \citet{SeppiComparatBulbul_2022A&A...665A..78S} to predict X-ray luminosities for each dark matter halo. 
The models agree with observations on a large scale, meaning that they sample the large-scale structure similarly. 

We find excellent agreement between the model and the measurement of C1xG1075 at small separations, suggesting that the quiescent fraction in the model is sensitive for galaxies with a $\log_{10}$ stellar mass greater than 10.75 and in clusters. 
At face value, this fraction is 90\% in the inner parts of the halos \citep[see Fig. 11][]{AungNagaiKlypin_2023MNRAS.519.1648A}. 
For C0xG1025, which samples lower stellar masses and lower redshifts, we find slight discrepancies at the $1\sigma$ level. 
To match the observations, the model must have a higher fraction of quiescent galaxies in the bin 10.25 to 10.75. 
The model quiescent fraction for such galaxies of about 80\% in the center \citep[see Fig. 11][]{AungNagaiKlypin_2023MNRAS.519.1648A} may need to be increased. 
The observed cross-correlation can thus serve as a benchmark to further constrain a model of the quiescent fraction as a function of cluster-centric radius.

In addition, we compare the measurements with predictions from the latest version of the L-Galaxies semi-analytical model \citep{AyromlouKauffmannYates_2021MNRAS.505..492A}. This model runs on subhalo merger trees from the Millennium simulation (box size $\sim$ 500 Mpc/$h$; \citealt{Springel2005}) and incorporates key physical processes, including gas accretion into dark matter halos, gas cooling, star formation, stellar feedback, chemical enrichment, and supermassive black hole-related mechanisms such as seeding, growth, and AGN feedback, as well as environmental effects. We use the L-Galaxies snapshot at $z \sim 0.2$, corresponding to the midpoint of our redshift bins.

Overall, L-Galaxies reproduces the 1-halo term of the cross-correlation function for the complete galaxy population in the C1xG1075 sample combination quite well. 
However, the model appears to over-predict the star-forming galaxy population when splitting galaxies into star-forming and quenched populations. It is important to note that the definition of quenching can affect this classification, as demonstrated by \cite{VaniAyromlouKauffmann_2025MNRAS.536..777V}, who compared quenched galaxy selections based on sSFR thresholds versus color-based criteria (e.g., rest-frame NUV-r and r-J colors).

{A more detailed comparison with predictions from large volume hydro-dynamical simulations \citep[e.g.][]{PakmorSpringelColes_2023MNRAS.524.2539P, SchayeKugelSchaller_2023MNRAS.526.4978S,NelsonPillepichAyromlou_2024A&A...686A.157N} should give us further insights into the possible relation between quenching and the thermodynamics of the hot gas in clusters \citep[e.g.][]{LimTacchellaMaiolino_2025arXiv250402027L}.} 

\subsection{Angular cross-correlation between clusters and galaxy split as a function of cluster redshift and galaxy color}
\label{subsec:measurements:colors}

\begin{figure*}
\centering
\includegraphics[width=0.95\columnwidth]{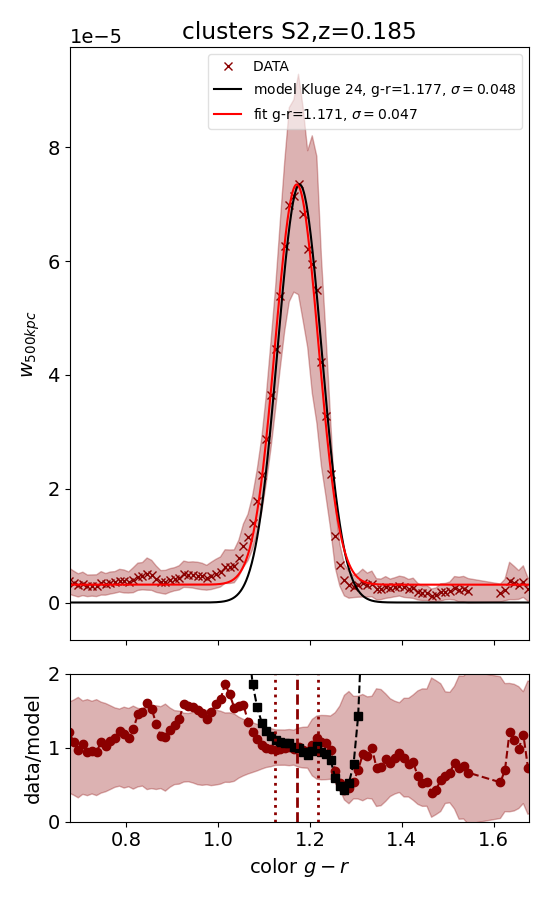}
\includegraphics[width=0.95\columnwidth]{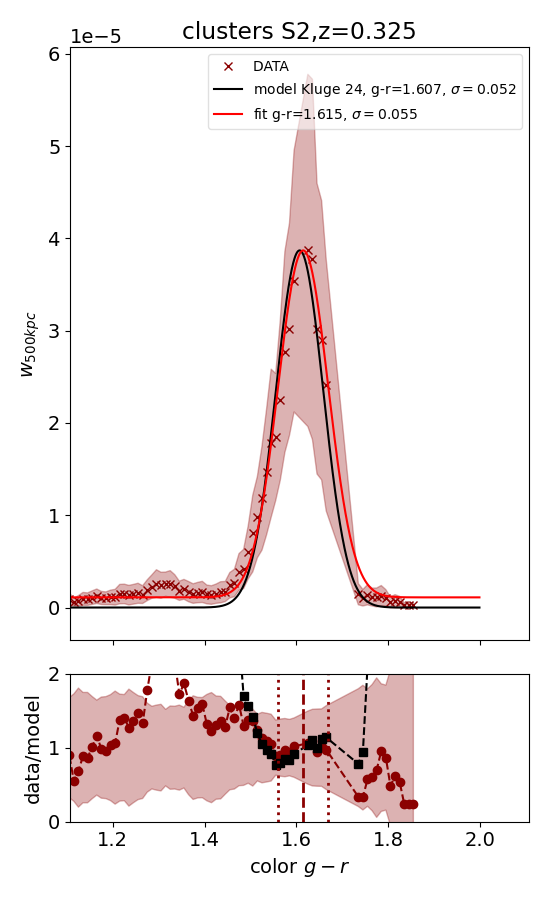}
\caption{Measured integral of the cross-correlation function up to 500 kpc as a function of color at redshift 0.195 (left column) and 0.325 (right column). 
We fit a Gaussian model to the data points (red). It compares well to the model from \citet{KlugeComparatLiu_2024A&A...688A.210K} in black.}
\label{fig:xcorr:fun:color}
\end{figure*}

We extend the cross-correlation analysis from the previous section by further splitting the samples as a function of galaxy color, here $g-r$ as measured in the legacy survey \citep[with DECam][]{FlaugherDiehlHonscheid_2015AJ....150..150F} and cluster redshift as measured by \citet{KlugeComparatLiu_2024A&A...688A.210K} using redmapper \citep{RykoffRozoBusha_2014ApJ...785..104R} to retrieve information about the galaxy red-sequence as a function of redshift. 

The accuracy of the galaxy photometric redshifts (0.05) is insufficient to cleanly separate the galaxy sample into fine and accurate redshift bins. 
Thus, we consider the complete galaxy sample described above (named BGS), selected to be brighter than $r_{AB}<19.5$ regardless of redshift, see Sec. \ref{subsec:data:galaxies}. 
The BGS sample is then split as a function of its observed $g-r$ color in 44 bins of 0.05 mag width in the range -0.2 to 2.
Since the complete galaxy population extends to redshift 0.4, we consider the C2 cluster samples from \citet{SeppiComparatGhirardini_2024A&A...686A.196S} selected with $0.1<z<0.4$, and $\log_{10}(L_X\; [\mathrm{erg/s}])>43.4$. 
The accuracy of the cluster photometric redshift in that range is about $\sim$0.005 \citep{KlugeComparatLiu_2024A&A...688A.210K}. 
We split the cluster sample using its redshifts in 30 bins of 0.01 width. 
The lowest and highest redshift bins of the cluster and random samples may suffer from boundary effects; therefore, we only consider the bins well within the sample in the range $0.15<z<0.35$. 
We then measure the cross-correlation between the 44 galaxy and 20 cluster samples, a total of  20$\times$44=880 cross-correlations.
For the complete BGS galaxy sample, we do not have a set of random points, {so we cannot use the \citet{LandySzalay_1993ApJ...412...64L} optimal estimator for the cross-correlation. We resort to a stacking technique by counting pairs of clusters and galaxies as a function of their separation and normalizing it by the equivalent counts centered at random positions (we have a random catalog for that cluster sample). 
The stacking is equivalent to the} \citet{DavisPeebles_1983ApJ...267..465D} estimator of the angular cross-correlation using galaxy-cluster pair counts that only needs a random sample for the cluster sample. 
\begin{equation}
\label{eq:xcorr:DP83}
w_{DP83}(\theta) = \frac{C\times G}{R\times G}\frac{N_R}{N_C} -1,    \end{equation}
where $C$ represents the clusters ($N_C$ their number), $G$ the galaxies, $R$ the cluster randoms  ($N_R$ their number). 

This estimator accurately estimates the clustering but will overestimate the variance on large scales (beyond a few arc minutes) compared to the \citet{LandySzalay_1993ApJ...412...64L} estimator {; see details of estimator comparisons in e.g. \citet{Pons-BorderiaMartinezStoyan_1999ApJ...523..480P,KerscherSzapudiSzalay_2000ApJ...535L..13K}}. 
However, since we will focus here on the 1-halo term and its integrated signal (up to 500kpc, corresponding to 2.5 arc minutes at z=0.2), this estimator is well-suited (but it would not be appropriate to estimate the large-scale clustering and its uncertainties as done in the previous section). 

We convert the angles into proper distances at the mean redshift of each cluster sample.
Most of the correlation functions are consistent with zero, with no signal. Some of them show a strong signal. 
To illustrate the strength of the correlation in the 1-halo term, we integrate all correlation functions up to 500kpc (proper distance) to obtain 
\begin{equation}
w_{500kpc} = \int^{500kpc}_0 w(\theta) d\theta   
\end{equation}
A 500kpc distance is well inside the 1-halo term of the clusters considered, and away from the 2-halo term to capture the extra power in clustering due to galaxies falling into the cluster potential well (see previous section and figures). 
The resulting values of $w_{500kpc}$ are shown in Fig. \ref{fig:xcorr:fun:color} left (right) panels as a function of color for two redshift values $z=$0.185 ($z=$0.325). 
 
We find that at fixed redshift, the integral of the cross-correlation captures the existence of the red sequence without any a priori assumptions about (i) the redshift of these galaxies, (ii) their possible membership in the clusters, (iii) stellar population synthesis models. 
Indeed, when compared to the model of \citet{KlugeComparatLiu_2024A&A...688A.210K} based on EZgal \citep{ManconeGonzalez_2012PASP..124..606M}, we find excellent agreement between the color at which this integral peaks (at a given redshift) and the color of the red sequence (Fig. \ref{fig:xcorr:fun:color}). 
We fit a Gaussian to the observations (see Fig. \ref{fig:xcorr:fun:color}) to retrieve the color value at which the correlation peaks and its spread, assuming it is Gaussian. 
The residuals show that the Gaussian approximation is fair to capture most of the observed curve. 
We repeat the fits on each redshift slice as a function of color. 
Table \ref{tab:RS:fit} gives the parameters obtained for each fit. 
They are very close to the model of \citet{KlugeComparatLiu_2024A&A...688A.210K}, but are derived independently. 
We measure accurately the scatter of the red sequence as a function of redshift. 

\begin{table}
    \centering
    \caption{Best-fit $g-r$ red-sequence value and its scatter as a function of redshift.}
    \begin{tabular}{ccc}
        \hline
       redshift & $g-r$ & scatter \\
       \hline
0.156 & 1.0618 $\pm$ 0.0007 & 0.0461 $\pm$ 0.0007 \\
0.165 & 1.0981 $\pm$ 0.0006 & 0.045 $\pm$ 0.0006 \\
0.175 & 1.1364 $\pm$ 0.0009 & 0.045 $\pm$ 0.001 \\
0.185 & 1.1712 $\pm$ 0.0005 & 0.0468 $\pm$ 0.0005 \\
0.195 & 1.2036 $\pm$ 0.0007 & 0.044 $\pm$ 0.0007 \\
0.205 & 1.2488 $\pm$ 0.0009 & 0.0438 $\pm$ 0.0009 \\
0.215 & 1.2837 $\pm$ 0.0008 & 0.0466 $\pm$ 0.0008 \\
0.225 & 1.3194 $\pm$ 0.0007 & 0.0495 $\pm$ 0.0007 \\
0.235 & 1.3498 $\pm$ 0.0008 & 0.0518 $\pm$ 0.0008 \\
0.245 & 1.3832 $\pm$ 0.0009 & 0.0556 $\pm$ 0.001 \\
0.255 & 1.4114 $\pm$ 0.0011 & 0.0576 $\pm$ 0.0011 \\
0.266 & 1.4409 $\pm$ 0.0009 & 0.0482 $\pm$ 0.0011 \\
0.275 & 1.4752 $\pm$ 0.001 & 0.059 $\pm$ 0.001 \\
0.285 & 1.5062 $\pm$ 0.0007 & 0.0546 $\pm$ 0.0008 \\
0.295 & 1.5373 $\pm$ 0.0008 & 0.0555 $\pm$ 0.0009 \\
0.305 & 1.5566 $\pm$ 0.0012 & 0.0552 $\pm$ 0.0013 \\
0.315 & 1.5766 $\pm$ 0.0009 & 0.0587 $\pm$ 0.001 \\
0.325 & 1.6148 $\pm$ 0.0008 & 0.0548 $\pm$ 0.0008 \\
0.334 & 1.6405 $\pm$ 0.0014 & 0.0539 $\pm$ 0.0014 \\
0.344 & 1.6762 $\pm$ 0.001 & 0.0417 $\pm$ 0.0011 \\
       \hline
    \end{tabular}
    \label{tab:RS:fit}
\end{table}

\begin{figure}
\centering
\includegraphics[width=0.95\columnwidth]{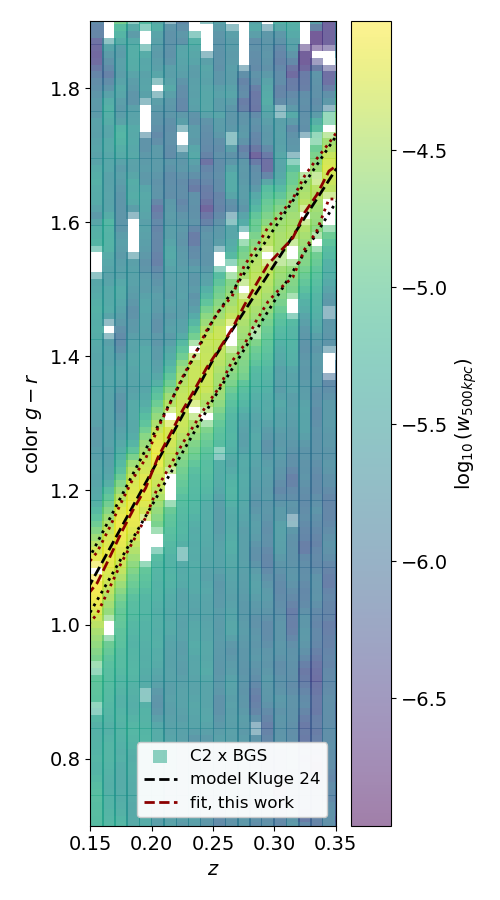}
\caption{g-r color vs. redshift colored with the integral of the cross-correlation function up to 500 kpc. Each colored symbol comes from a cross-correlation function. The dashed red lines link the models done on each redshift slice. It is in excellent agreement with the red-sequence (black lines) model from \citet{KlugeComparatLiu_2024A&A...688A.210K}. We recover the red sequence and its scatter accurately.}
\label{fig:redseq}
\end{figure}

We show the values of $w_{500kpc}$ obtained for all correlation functions in the color-redshift plane in Fig. \ref{fig:redseq}. 
With the fits described above, we locate our best-fit red-sequence (red dashed) in this plane. The agreement is nearly perfect with the model of \citet{KlugeComparatLiu_2024A&A...688A.210K}, shown in black. 

\section{Discussion and outlook}
\label{sec:discussion}

In this study, we measure the cross-correlation between volume-limited samples of galaxies and X-ray-selected clusters with unprecedented signal-to-noise (Sec. \ref{subsec:measurements:rsbc}). Thanks to this high level of significance, we can finely split the samples over a grid of redshift and color to measure the amplitude of its one-halo term via the integral of the cluster-galaxy cross-correlation within 500kpc (Sec. \ref{subsec:measurements:colors}). 

We report two main findings. 
Firstly, we find that the cross-correlation between clusters and galaxies split into quiescent and star-forming galaxies contains key information about the infall of blue cloud galaxies, feedback, and quenching in a high mass environment (Fig. \ref{fig:wprp:C1xG1075} and \ref{fig:wprp:C0xG1025}). 
The results are in excellent quantitative agreement for stellar masses above $\log_{10}(M^*[M_\odot])>10.75$ with the empirical galaxy evolution model of \citet{BehrooziWechslerHearin_2019MNRAS.488.3143B, AungNagaiKlypin_2023MNRAS.519.1648A}. 
The measurements also suggest that in the stellar mass range 10.25$<\log_{10}(M^*[M_\odot])<$10.75, the red fraction may be slightly underestimated by the model (Fig. \ref{fig:CxG:UM}). 
The results are in fair qualitative agreement with the LGAL model \citep{AyromlouKauffmannYates_2021MNRAS.505..492A}.

Secondly, we find that the integral of the cross-correlation ($w_{500kpc}$) between clusters (split in redshifts) and galaxies (split in colors) enables a direct measurement of the red sequence and its scatter in galaxy clusters (Fig. \ref{fig:xcorr:fun:color} and \ref{fig:redseq}). This measurement is independent of models of galaxy evolution or our ability to classify galaxies as cluster members. 
This method constitutes an alternative to the calibration of red-sequence models from \citep{RykoffRozoBusha_2014ApJ...785..104R}. 
Similarly to the \citet{RykoffRozoBusha_2014ApJ...785..104R} method, it requires a well-understood cluster sample, with accurate photometric redshifts, which is here provided by the X-ray selected eRASS1 catalogue from \citet{SeppiComparatGhirardini_2024A&A...686A.196S} and its volume-limited samples. 
As for the galaxies, this method requires a complete (clean) flux-limited galaxy sample (no redshifts required, only colors and sky positions), here provided by the LS10 sample of \citet{ComparatMerloniPonti_2025A&A...697A.173C}. 
The \citet{RykoffRozoBusha_2014ApJ...785..104R} method requires, in addition, a stellar population synthesis model \citep[e.g.][]{BruzualCharlot_2003MNRAS.344.1000B}, and a set of known galaxy members (with spectroscopic redshifts) that are in the red-sequence \citep[e.g.][]{ClercKirkpatrickFinoguenov_2020MNRAS.497.3976C}. 

\subsection*{Discussion}

With these new measurements, assuming models of how galaxies populate clusters, one can constrain the relative abundance of red and blue galaxies in clusters and possibly how infall, feedback, and quenching are related \citep[see e.g.][]{DeLuciaBlaizot_2007MNRAS.375....2D, AyromlouKauffmannAnand_2023MNRAS.519..1913A, AungNagaiKlypin_2023MNRAS.519.1648A, LehleNelsonPillepich_2024A&A...687A.129L}. 

Alternatively, one could constrain the average profile and concentration of red-sequence and blue-cloud galaxies in galaxy clusters with a halo model. 
The quality of the summary statistic measured here would enable us to go beyond current constraints from \citet{HennigMohrZenteno_2017MNRAS.467.4015H, Nishizawa2018PASJ...70S..24N}. 
We may constrain where the transition from the field to cluster galaxies occurs and at which radius ram-pressure stripping becomes essential. 
 
\subsection*{Outlook}

Splitting the cluster samples as a function of halo mass may be interesting. One could expect the situation in low-mass clusters or groups to differ significantly from that of massive clusters. This may become possible with upcoming cluster samples from eROSITA, Euclid, LSST, and CMB-S4. Indeed, the ram pressure depends on the square of the infall velocity \citep[see recent review from ][]{BoselliFossatiSun_2022A&ARv..30....3B}.

A similar analysis using SZ detected cluster \citep[e.g., ACT \& Planck][]{NaessAiolaAustermann_2020JCAP...12..046N,NaessGuanDuivenvoorden_2025arXiv250314451N} or in the future CMB-S4 \citep{CMB-S4_AbazajianAdsheadAhmed_2016arXiv161002743A} and this very same set of galaxies would be complementary but is outside the scope of this study.

In the future, with Euclid \citep{EuclidCollaborationMellierAbdurro'uf_2025A&A...697A...1E} and LSST \citep{IvezicKahnTyson_2019ApJ...873..111I}, and their upcoming photometric galaxy samples together with 4MOST and DESI spectroscopic samples \citep{deJongAgertzBerbel_2019Msngr.175....3D, DESICollaborationAghamousaAguilar_2016arXiv161100036D, DESICollaborationAghamousaAguilar_2016arXiv161100037D}, we expect the summary statistics shown here to see a significant increase in its signal to noise ratio, in particular at small separations where lowering the stellar mass threshold (by increasing the magnitude limit) is key. 

The cross-correlation between clusters selected in X-ray, SZ, optical, infrared \citep[e.g.][]{BulbulLiuKluge_2024A&A...685A.106B,NaessGuanDuivenvoorden_2025arXiv250314451N,RykoffRozoBusha_2014ApJ...785..104R,EuclidCollaborationBhargavaBenoist_2025arXiv250319196E} and a fixed sample galaxy (that is as pure and complete as possible) will contain essential information about the cluster selection function and how they relate to the large-scale structure \citep[e.g.][]{ClercComparatSeppi_2024A&A...687A.238C}. 
A model for a cluster selection function that reproduces the observed cross-correlation with an external data set ought to be more trustworthy.

\section*{Data availability} 

The auto and cross-correlations measured and presented in Sec. \ref{subsec:measurements:rsbc} are available on 
\href{https://zenodo.org/records/15806800}{Zenodo}. 

\bibliographystyle{aa}
\bibliography{references}

\begin{thebibliography}{103}
\expandafter\ifx\csname natexlab\endcsname\relax\def\natexlab#1{#1}\fi

\bibitem[{{Abazajian} {et~al.}(2016){Abazajian}, {Adshead}, {Ahmed}, {Allen},
  {Alonso}, {Arnold}, {Baccigalupi}, {Bartlett}, {Battaglia}, {Benson}, \&
  et~al.}]{CMB-S4_AbazajianAdsheadAhmed_2016arXiv161002743A}
{Abazajian}, K.~N., {Adshead}, P., {Ahmed}, Z., {et~al.} 2016, arXiv e-prints,
  arXiv:1610.02743

\bibitem[{{Adhikari} {et~al.}(2021){Adhikari}, {Shin}, {Jain}, {Hilton},
  {Baxter}, {Chang}, {Wechsler}, {Battaglia}, {Bond}, {Bocquet}, \&
  et~al.}]{AdhikariShinJain_2021ApJ...923...37A}
{Adhikari}, S., {Shin}, T.-h., {Jain}, B., {et~al.} 2021, \apj, 923, 37

\bibitem[{{Amodeo} {et~al.}(2021){Amodeo}, {Battaglia}, {Schaan}, {Ferraro},
  {Moser}, {Aiola}, {Austermann}, {Beall}, {Bean}, {Becker}, \&
  et~al.}]{AmodeoBattagliaSchaan_2021PhRvD.103f3514A}
{Amodeo}, S., {Battaglia}, N., {Schaan}, E., {et~al.} 2021, \prd, 103, 063514

\bibitem[{{Aung} {et~al.}(2023){Aung}, {Nagai}, {Klypin}, {Behroozi},
  {Abdullah}, {Ishiyama}, {Prada}, {P{\'e}rez}, {L{\'o}pez Cacheiro}, \&
  {Ruedas}}]{AungNagaiKlypin_2023MNRAS.519.1648A}
{Aung}, H., {Nagai}, D., {Klypin}, A., {et~al.} 2023, \mnras, 519, 1648

\bibitem[{{Ayromlou} {et~al.}(2023){Ayromlou}, {Kauffmann}, {Anand}, \&
  {White}}]{AyromlouKauffmannAnand_2023MNRAS.519..1913A}
{Ayromlou}, M., {Kauffmann}, G., {Anand}, A., \& {White}, S. D.~M. 2023,
  \mnras, 519, 1913

\bibitem[{{Ayromlou} {et~al.}(2021){Ayromlou}, {Kauffmann}, {Yates}, {Nelson},
  \& {White}}]{AyromlouKauffmannYates_2021MNRAS.505..492A}
{Ayromlou}, M., {Kauffmann}, G., {Yates}, R.~M., {Nelson}, D., \& {White}, S.
  D.~M. 2021, \mnras, 505, 492

\bibitem[{{Banerjee} \& {Abel}(2021)}]{BanerjeeAbel_2021MNRAS.504.2911B}
{Banerjee}, A. \& {Abel}, T. 2021, \mnras, 504, 2911

\bibitem[{{Baxter} {et~al.}(2017){Baxter}, {Chang}, {Jain}, {Adhikari},
  {Dalal}, {Kravtsov}, {More}, {Rozo}, {Rykoff}, \&
  {Sheth}}]{BaxterChangJain_2017ApJ...841...18B}
{Baxter}, E., {Chang}, C., {Jain}, B., {et~al.} 2017, \apj, 841, 18

\bibitem[{{Behroozi} {et~al.}(2019){Behroozi}, {Wechsler}, {Hearin}, \&
  {Conroy}}]{BehrooziWechslerHearin_2019MNRAS.488.3143B}
{Behroozi}, P., {Wechsler}, R.~H., {Hearin}, A.~P., \& {Conroy}, C. 2019,
  \mnras, 488, 3143

\bibitem[{{Bell} {et~al.}(2004){Bell}, {Wolf}, {Meisenheimer}, {Rix}, {Borch},
  {Dye}, {Kleinheinrich}, {Wisotzki}, \&
  {McIntosh}}]{BellWolfMeisenheimer_2004ApJ...608..752B}
{Bell}, E.~F., {Wolf}, C., {Meisenheimer}, K., {et~al.} 2004, \apj, 608, 752

\bibitem[{{Bianconi} {et~al.}(2020){Bianconi}, {Smith}, {Haines}, {McGee},
  {Finoguenov}, \& {Egami}}]{BianconiSmithHaines_2020MNRAS.492.4599B}
{Bianconi}, M., {Smith}, G.~P., {Haines}, C.~P., {et~al.} 2020, \mnras, 492,
  4599

\bibitem[{{B{\"o}hringer} {et~al.}(2004){B{\"o}hringer}, {Schuecker}, {Guzzo},
  {Collins}, {Voges}, {Cruddace}, {Ortiz-Gil}, {Chincarini}, {De Grandi},
  {Edge}, \& et~al.}]{BohringerSchueckerGuzzo_2004A&A...425..367B}
{B{\"o}hringer}, H., {Schuecker}, P., {Guzzo}, L., {et~al.} 2004, \aap, 425,
  367

\bibitem[{{Boselli} {et~al.}(2022){Boselli}, {Fossati}, \&
  {Sun}}]{BoselliFossatiSun_2022A&ARv..30....3B}
{Boselli}, A., {Fossati}, M., \& {Sun}, M. 2022, \aapr, 30, 3

\bibitem[{{Bruzual} \& {Charlot}(2003)}]{BruzualCharlot_2003MNRAS.344.1000B}
{Bruzual}, G. \& {Charlot}, S. 2003, \mnras, 344, 1000

\bibitem[{{Bulbul} {et~al.}(2024){Bulbul}, {Liu}, {Kluge}, {Zhang}, {Sanders},
  {Bahar}, {Ghirardini}, {Artis}, {Seppi}, {Garrel}, \&
  et~al.}]{BulbulLiuKluge_2024A&A...685A.106B}
{Bulbul}, E., {Liu}, A., {Kluge}, M., {et~al.} 2024, \aap, 685, A106

\bibitem[{{Clerc} {et~al.}(2024){Clerc}, {Comparat}, {Seppi}, {Artis}, {Bahar},
  {Balzer}, {Bulbul}, {Dauser}, {Garrel}, {Ghirardini}, \&
  et~al.}]{ClercComparatSeppi_2024A&A...687A.238C}
{Clerc}, N., {Comparat}, J., {Seppi}, R., {et~al.} 2024, \aap, 687, A238

\bibitem[{{Clerc} {et~al.}(2020){Clerc}, {Kirkpatrick}, {Finoguenov},
  {Capasso}, {Comparat}, {Damsted}, {Furnell}, {Kukkola}, {Ider Chitham},
  {Merloni}, \& et~al.}]{ClercKirkpatrickFinoguenov_2020MNRAS.497.3976C}
{Clerc}, N., {Kirkpatrick}, C.~C., {Finoguenov}, A., {et~al.} 2020, \mnras,
  497, 3976

\bibitem[{{Comparat} {et~al.}(2020){Comparat}, {Eckert}, {Finoguenov},
  {Schmidt}, {Sanders}, {Nagai}, {Lau}, {K{\"a}}, {fer}, {Pacaud}, \&
  et~al.}]{ComparatEckertFinoguenov_2020OJAp....3E..13C}
{Comparat}, J., {Eckert}, D., {Finoguenov}, A., {et~al.} 2020, The Open Journal
  of Astrophysics, 3, 13

\bibitem[{{Comparat} {et~al.}(2025){Comparat}, {Merloni}, {Ponti}, {Shreeram},
  {Zhang}, {Reiprich}, {Liu}, {Seppi}, {Zhang}, {Clerc}, \&
  et~al.}]{ComparatMerloniPonti_2025A&A...697A.173C}
{Comparat}, J., {Merloni}, A., {Ponti}, G., {et~al.} 2025, \aap, 697, A173

\bibitem[{{Comparat} {et~al.}(2022){Comparat}, {Truong}, {Merloni},
  {Pillepich}, {Ponti}, {Driver}, {Bellstedt}, {Liske}, {Aird}, {Br{\"u}ggen},
  \& et~al.}]{ComparatTruongMerloni_2022A&A...666A.156C}
{Comparat}, J., {Truong}, N., {Merloni}, A., {et~al.} 2022, \aap, 666, A156

\bibitem[{{Cooray} \& {Sheth}(2002)}]{CooraySheth_2002PhR...372....1C}
{Cooray}, A. \& {Sheth}, R. 2002, \physrep, 372, 1

\bibitem[{{Croft} {et~al.}(1999){Croft}, {Dalton}, \&
  {Efstathiou}}]{CroftDaltonEfstathiou_1999MNRAS.305..547C}
{Croft}, R. A.~C., {Dalton}, G.~B., \& {Efstathiou}, G. 1999, \mnras, 305, 547

\bibitem[{{Dalton} {et~al.}(1997){Dalton}, {Maddox}, {Sutherland}, \&
  {Efstathiou}}]{DaltonMaddoxSutherland_1997MNRAS.289..263D}
{Dalton}, G.~B., {Maddox}, S.~J., {Sutherland}, W.~J., \& {Efstathiou}, G.
  1997, \mnras, 289, 263

\bibitem[{{Das} {et~al.}(2023){Das}, {Chiang}, \&
  {Mathur}}]{DasChiangMathur_2023ApJ...951..125D}
{Das}, S., {Chiang}, Y.-K., \& {Mathur}, S. 2023, \apj, 951, 125

\bibitem[{{Davies} {et~al.}(2019){Davies}, {Crain}, {McCarthy}, {Oppenheimer},
  {Schaye}, {Schaller}, \&
  {McAlpine}}]{DaviesCrainMcCarthy_2019MNRAS.485.3783D}
{Davies}, J.~J., {Crain}, R.~A., {McCarthy}, I.~G., {et~al.} 2019, \mnras, 485,
  3783

\bibitem[{{Davis} \& {Peebles}(1983)}]{DavisPeebles_1983ApJ...267..465D}
{Davis}, M. \& {Peebles}, P.~J.~E. 1983, \apj, 267, 465

\bibitem[{{de Jong} {et~al.}(2019){de Jong}, {Agertz}, {Berbel}, {Aird},
  {Alexander}, {Amarsi}, {Anders}, {Andrae}, {Ansarinejad}, {Ansorge}, \&
  et~al.}]{deJongAgertzBerbel_2019Msngr.175....3D}
{de Jong}, R.~S., {Agertz}, O., {Berbel}, A.~A., {et~al.} 2019, The Messenger,
  175, 3

\bibitem[{{De Lucia} \& {Blaizot}(2007)}]{DeLuciaBlaizot_2007MNRAS.375....2D}
{De Lucia}, G. \& {Blaizot}, J. 2007, \mnras, 375, 2

\bibitem[{{DESI Collaboration} {et~al.}(2016{\natexlab{a}}){DESI
  Collaboration}, {Aghamousa}, {Aguilar}, {Ahlen}, {Alam}, {Allen}, {Allende
  Prieto}, {Annis}, {Bailey}, {Balland}, \&
  et~al.}]{DESICollaborationAghamousaAguilar_2016arXiv161100036D}
{DESI Collaboration}, {Aghamousa}, A., {Aguilar}, J., {et~al.}
  2016{\natexlab{a}}, arXiv e-prints, arXiv:1611.00036

\bibitem[{{DESI Collaboration} {et~al.}(2016{\natexlab{b}}){DESI
  Collaboration}, {Aghamousa}, {Aguilar}, {Ahlen}, {Alam}, {Allen}, {Allende
  Prieto}, {Annis}, {Bailey}, {Balland}, \&
  et~al.}]{DESICollaborationAghamousaAguilar_2016arXiv161100037D}
{DESI Collaboration}, {Aghamousa}, A., {Aguilar}, J., {et~al.}
  2016{\natexlab{b}}, arXiv e-prints, arXiv:1611.00037

\bibitem[{{Dev} {et~al.}(2024){Dev}, {Driver}, {Meyer}, {Robotham},
  {Obreschkow}, {Popesso}, \& {Comparat}}]{DevDriverMeyer_2024MNRAS.535.2357D}
{Dev}, A., {Driver}, S.~P., {Meyer}, M., {et~al.} 2024, \mnras, 535, 2357

\bibitem[{{Dey} {et~al.}(2019){Dey}, {Schlegel}, {Lang}, {Blum}, {Burleigh},
  {Fan}, {Findlay}, {Finkbeiner}, {Herrera}, {Juneau}, \&
  et~al.}]{DeySchlegelLang_2019AJ....157..168D}
{Dey}, A., {Schlegel}, D.~J., {Lang}, D., {et~al.} 2019, \aj, 157, 168

\bibitem[{{Euclid Collaboration} {et~al.}(2025){Euclid Collaboration},
  {Mellier}, {Abdurro'uf}, {Acevedo Barroso}, {Ach{\'u}carro}, {Adamek},
  {Adam}, {Addison}, {Aghanim}, {Aguena}, \&
  et~al.}]{EuclidCollaborationMellierAbdurro'uf_2025A&A...697A...1E}
{Euclid Collaboration}, {Mellier}, Y., {Abdurro'uf}, {et~al.} 2025, \aap, 697,
  A1

\bibitem[{{Euclid Collaboration: Bhargava} {et~al.}(2025){Euclid Collaboration:
  Bhargava}, {Benoist}, {Gonzalez}, {Maturi}, {Melin}, {Stanford}, {Munari},
  {Vannier}, {Murray}, \&
  et~al.}]{EuclidCollaborationBhargavaBenoist_2025arXiv250319196E}
{Euclid Collaboration: Bhargava}, S., {Benoist}, C., {Gonzalez}, A.~H.,
  {et~al.} 2025, arXiv e-prints, arXiv:2503.19196

\bibitem[{{Faber} {et~al.}(2007){Faber}, {Willmer}, {Wolf}, {Koo}, {Weiner},
  {Newman}, {Im}, {Coil}, {Conroy}, {Cooper}, \&
  et~al.}]{FaberWillmerWolf_2007ApJ...665..265F}
{Faber}, S.~M., {Willmer}, C.~N.~A., {Wolf}, C., {et~al.} 2007, \apj, 665, 265

\bibitem[{{Falck} {et~al.}(2021){Falck}, {Wang}, {Jenkins}, {Lemson},
  {Medvedev}, {Neyrinck}, \& {Szalay}}]{FalckWangJenkins_2021MNRAS.506.2659F}
{Falck}, B., {Wang}, J., {Jenkins}, A., {et~al.} 2021, \mnras, 506, 2659

\bibitem[{{Fedeli} {et~al.}(2011){Fedeli}, {Carbone}, {Moscardini}, \&
  {Cimatti}}]{FedeliCarboneMoscardini_2011MNRAS.414.1545F}
{Fedeli}, C., {Carbone}, C., {Moscardini}, L., \& {Cimatti}, A. 2011, \mnras,
  414, 1545

\bibitem[{{Flaugher} {et~al.}(2015){Flaugher}, {Diehl}, {Honscheid}, {Abbott},
  {Alvarez}, {Angstadt}, {Annis}, {Antonik}, {Ballester}, {Beaufore}, \&
  et~al.}]{FlaugherDiehlHonscheid_2015AJ....150..150F}
{Flaugher}, B., {Diehl}, H.~T., {Honscheid}, K., {et~al.} 2015, \aj, 150, 150

\bibitem[{{Gladders} \& {Yee}(2000)}]{GladdersYee_2000AJ....120.2148G}
{Gladders}, M.~D. \& {Yee}, H.~K.~C. 2000, \aj, 120, 2148

\bibitem[{{Grayson} {et~al.}(2023){Grayson}, {Scannapieco}, \&
  {Dav{\'e}}}]{GraysonScannapiecoDave_2023ApJ...957...17G}
{Grayson}, S., {Scannapieco}, E., \& {Dav{\'e}}, R. 2023, \apj, 957, 17

\bibitem[{{Hahn} {et~al.}(2023){Hahn}, {Wilson}, {Ruiz-Macias}, {Cole},
  {Weinberg}, {Moustakas}, {Kremin}, {Tinker}, {Smith}, {Wechsler}, \&
  et~al.}]{HahnWilsonRuiz-Macias_2023AJ....165..253H}
{Hahn}, C., {Wilson}, M.~J., {Ruiz-Macias}, O., {et~al.} 2023, \aj, 165, 253

\bibitem[{{Hennig} {et~al.}(2017){Hennig}, {Mohr}, {Zenteno}, {Desai},
  {Dietrich}, {Bocquet}, {Strazzullo}, {Saro}, {Abbott}, {Abdalla}, \&
  et~al.}]{HennigMohrZenteno_2017MNRAS.467.4015H}
{Hennig}, C., {Mohr}, J.~J., {Zenteno}, A., {et~al.} 2017, \mnras, 467, 4015

\bibitem[{{H{\"u}tsi} \& {Lahav}(2008)}]{HutsiLahav_2008A&A...492..355H}
{H{\"u}tsi}, G. \& {Lahav}, O. 2008, \aap, 492, 355

\bibitem[{{Ibitoye} {et~al.}(2022){Ibitoye}, {Tramonte}, {Ma}, \&
  {Dai}}]{IbitoyeTramonteMa_2022ApJ...935...18I}
{Ibitoye}, A., {Tramonte}, D., {Ma}, Y.-Z., \& {Dai}, W.-M. 2022, \apj, 935, 18

\bibitem[{{Ishiyama} {et~al.}(2021){Ishiyama}, {Prada}, {Klypin}, {Sinha},
  {Metcalf}, {Jullo}, {Altieri}, {Cora}, {Croton}, {de la Torre}, \&
  et~al.}]{IshiyamaPradaKlypin_2021MNRAS.506.4210I}
{Ishiyama}, T., {Prada}, F., {Klypin}, A.~A., {et~al.} 2021, \mnras, 506, 4210

\bibitem[{{Ivezi{\'c}} {et~al.}(2019){Ivezi{\'c}}, {Kahn}, {Tyson}, {Abel},
  {Acosta}, {Allsman}, {Alonso}, {AlSayyad}, {Anderson}, {Andrew}, \&
  et~al.}]{IvezicKahnTyson_2019ApJ...873..111I}
{Ivezi{\'c}}, {\v{Z}}., {Kahn}, S.~M., {Tyson}, J.~A., {et~al.} 2019, \apj,
  873, 111

\bibitem[{{Jarrett} {et~al.}(2000){Jarrett}, {Chester}, {Cutri}, {Schneider},
  {Skrutskie}, \& {Huchra}}]{JarrettChesterCutri_2000AJ....119.2498J}
{Jarrett}, T.~H., {Chester}, T., {Cutri}, R., {et~al.} 2000, \aj, 119, 2498

\bibitem[{{Kerscher} {et~al.}(2000){Kerscher}, {Szapudi}, \&
  {Szalay}}]{KerscherSzapudiSzalay_2000ApJ...535L..13K}
{Kerscher}, M., {Szapudi}, I., \& {Szalay}, A.~S. 2000, \apjl, 535, L13

\bibitem[{{Kluge} {et~al.}(2024){Kluge}, {Comparat}, {Liu}, {Balzer}, {Bulbul},
  {Ider Chitham}, {Ghirardini}, {Garrel}, {Bahar}, {Artis}, \&
  et~al.}]{KlugeComparatLiu_2024A&A...688A.210K}
{Kluge}, M., {Comparat}, J., {Liu}, A., {et~al.} 2024, \aap, 688, A210

\bibitem[{{Koester} {et~al.}(2007){Koester}, {McKay}, {Annis}, {Wechsler},
  {Evrard}, {Rozo}, {Bleem}, {Sheldon}, \&
  {Johnston}}]{KoesterMcKayAnnis_2007ApJ...660..221K}
{Koester}, B.~P., {McKay}, T.~A., {Annis}, J., {et~al.} 2007, \apj, 660, 221

\bibitem[{{Kukstas} {et~al.}(2020){Kukstas}, {McCarthy}, {Baldry}, \&
  {Font}}]{KukstasMcCarthyBaldry_2020MNRAS.496.2241K}
{Kukstas}, E., {McCarthy}, I.~G., {Baldry}, I.~K., \& {Font}, A.~S. 2020,
  \mnras, 496, 2241

\bibitem[{{La Posta} {et~al.}(2024){La Posta}, {Alonso}, {Chisari}, {Ferreira},
  \& {Garc{\'\i}a-Garc{\'\i}a}}]{LaPostaAlonsoChisari_2024arXiv241212081L}
{La Posta}, A., {Alonso}, D., {Chisari}, N.~E., {Ferreira}, T., \&
  {Garc{\'\i}a-Garc{\'\i}a}, C. 2024, arXiv e-prints, arXiv:2412.12081

\bibitem[{{Landy} \& {Szalay}(1993)}]{LandySzalay_1993ApJ...412...64L}
{Landy}, S.~D. \& {Szalay}, A.~S. 1993, \apj, 412, 64

\bibitem[{{Lau} {et~al.}(2024){Lau}, {Nagai}, {Bogd{\'a}n}, {Medlock},
  {Oppenheimer}, {Battaglia}, {Angl{\'e}s-Alc{\'a}zar}, {Genel}, {Ni}, \&
  {Villaescusa-Navarro}}]{LauNagaiBogdan_2024arXiv241204559L}
{Lau}, E.~T., {Nagai}, D., {Bogd{\'a}n}, {\'A}., {et~al.} 2024, arXiv e-prints,
  arXiv:2412.04559

\bibitem[{{Laureijs} {et~al.}(2011){Laureijs}, {Amiaux}, {Arduini},
  {Augu{\`e}res}, {Brinchmann}, {Cole}, {Cropper}, {Dabin}, {Duvet}, {Ealet},
  \& et~al.}]{LaureijsAmiauxArduini_2011arXiv1110.3193L}
{Laureijs}, R., {Amiaux}, J., {Arduini}, S., {et~al.} 2011, arXiv e-prints,
  arXiv:1110.3193

\bibitem[{{Lehle} {et~al.}(2024){Lehle}, {Nelson}, {Pillepich}, {Truong}, \&
  {Rohr}}]{LehleNelsonPillepich_2024A&A...687A.129L}
{Lehle}, K., {Nelson}, D., {Pillepich}, A., {Truong}, N., \& {Rohr}, E. 2024,
  \aap, 687, A129

\bibitem[{{Li} {et~al.}(2024){Li}, {Fang}, {Ge}, {Liu}, {He}, {Li}, {Nicastro},
  {Yang}, {Zhang}, \& {Zheng}}]{LiFangGe_2024ApJ...977L..40L}
{Li}, D., {Fang}, T., {Ge}, C., {et~al.} 2024, \apjl, 977, L40

\bibitem[{{Lilje} \& {Efstathiou}(1988)}]{LiljeEfstathiou_1988MNRAS.231..635L}
{Lilje}, P.~B. \& {Efstathiou}, G. 1988, \mnras, 231, 635

\bibitem[{{Lim} {et~al.}(2025){Lim}, {Tacchella}, {Maiolino}, {Schaye}, \&
  {Schaller}}]{LimTacchellaMaiolino_2025arXiv250402027L}
{Lim}, S., {Tacchella}, S., {Maiolino}, R., {Schaye}, J., \& {Schaller}, M.
  2025, arXiv e-prints, arXiv:2504.02027

\bibitem[{{Maddox} {et~al.}(1990){Maddox}, {Sutherland}, {Efstathiou}, \&
  {Loveday}}]{MaddoxSutherlandEfstathiou_1990MNRAS.243..692M}
{Maddox}, S.~J., {Sutherland}, W.~J., {Efstathiou}, G., \& {Loveday}, J. 1990,
  \mnras, 243, 692

\bibitem[{{Mancone} \& {Gonzalez}(2012)}]{ManconeGonzalez_2012PASP..124..606M}
{Mancone}, C.~L. \& {Gonzalez}, A.~H. 2012, \pasp, 124, 606

\bibitem[{{Merloni} {et~al.}(2024){Merloni}, {Lamer}, {Liu}, {Ramos-Ceja},
  {Brunner}, {Bulbul}, {Dennerl}, {Doroshenko}, {Freyberg}, {Friedrich}, \&
  et~al.}]{MerloniLamerLiu_2024A&A...682A..34M}
{Merloni}, A., {Lamer}, G., {Liu}, T., {et~al.} 2024, \aap, 682, A34

\bibitem[{{Merloni} {et~al.}(2012){Merloni}, {Predehl}, {Becker},
  {B{\"o}hringer}, {Boller}, {Brunner}, {Brusa}, {Dennerl}, {Freyberg},
  {Friedrich}, \& et~al.}]{MerloniPredehlBecker_2012arXiv1209.3114M}
{Merloni}, A., {Predehl}, P., {Becker}, W., {et~al.} 2012, arXiv e-prints,
  arXiv:1209.3114

\bibitem[{{More} {et~al.}(2015){More}, {Miyatake}, {Mandelbaum}, {Takada},
  {Spergel}, {Brownstein}, \&
  {Schneider}}]{MoreMiyatakeMandelbaum_2015ApJ...806....2M}
{More}, S., {Miyatake}, H., {Mandelbaum}, R., {et~al.} 2015, \apj, 806, 2

\bibitem[{{Moser} {et~al.}(2022){Moser}, {Battaglia}, {Nagai}, {Lau}, {Machado
  Poletti Valle}, {Villaescusa-Navarro}, {Amodeo}, {Angl{\'e}s-Alc{\'a}zar},
  {Bryan}, {Dave}, \& et~al.}]{MoserBattagliaNagai_2022ApJ...933..133M}
{Moser}, E., {Battaglia}, N., {Nagai}, D., {et~al.} 2022, \apj, 933, 133

\bibitem[{{Naess} {et~al.}(2020){Naess}, {Aiola}, {Austermann}, {Battaglia},
  {Beall}, {Becker}, {Bond}, {Calabrese}, {Choi}, {Cothard}, \&
  et~al.}]{NaessAiolaAustermann_2020JCAP...12..046N}
{Naess}, S., {Aiola}, S., {Austermann}, J.~E., {et~al.} 2020, \jcap, 2020, 046

\bibitem[{{Naess} {et~al.}(2025){Naess}, {Guan}, {Duivenvoorden},
  {Hasselfield}, {Wang}, {Abril-Cabezas}, {Addison}, {Ade}, {Aiola}, {Alford},
  \& et~al.}]{NaessGuanDuivenvoorden_2025arXiv250314451N}
{Naess}, S., {Guan}, Y., {Duivenvoorden}, A.~J., {et~al.} 2025, arXiv e-prints,
  arXiv:2503.14451

\bibitem[{{Nelan} {et~al.}(2005){Nelan}, {Smith}, {Hudson}, {Wegner}, {Lucey},
  {Moore}, {Quinney}, \& {Suntzeff}}]{NelanSmithHudson_2005ApJ...632..137N}
{Nelan}, J.~E., {Smith}, R.~J., {Hudson}, M.~J., {et~al.} 2005, \apj, 632, 137

\bibitem[{{Nelson} {et~al.}(2024){Nelson}, {Pillepich}, {Ayromlou}, {Lee},
  {Lehle}, {Rohr}, \& {Truong}}]{NelsonPillepichAyromlou_2024A&A...686A.157N}
{Nelson}, D., {Pillepich}, A., {Ayromlou}, M., {et~al.} 2024, \aap, 686, A157

\bibitem[{{Nishizawa} {et~al.}(2018){Nishizawa}, {Oguri}, {Oogi}, {More},
  {Nishimichi}, {Nagashima}, {Lin}, {Mandelbaum}, {Takada}, {Bahcall},
  {Coupon}, {Huang}, {Jian}, {Komiyama}, {Leauthaud}, {Lin}, {Miyatake},
  {Miyazaki}, \& {Tanaka}}]{Nishizawa2018PASJ...70S..24N}
{Nishizawa}, A.~J., {Oguri}, M., {Oogi}, T., {et~al.} 2018, \pasj, 70, S24

\bibitem[{{Oren} {et~al.}(2024){Oren}, {Sternberg}, {McKee}, {Faerman}, \&
  {Genel}}]{OrenSternbergMcKee_2024ApJ...974..291O}
{Oren}, Y., {Sternberg}, A., {McKee}, C.~F., {Faerman}, Y., \& {Genel}, S.
  2024, \apj, 974, 291

\bibitem[{{Paech} {et~al.}(2017){Paech}, {Hamaus}, {Hoyle}, {Costanzi},
  {Giannantonio}, {Hagstotz}, {Sauerwein}, \&
  {Weller}}]{PaechHamausHoyle_2017MNRAS.470.2566P}
{Paech}, K., {Hamaus}, N., {Hoyle}, B., {et~al.} 2017, \mnras, 470, 2566

\bibitem[{{Pakmor} {et~al.}(2023){Pakmor}, {Springel}, {Coles}, {Guillet},
  {Pfrommer}, {Bose}, {Barrera}, {Delgado}, {Ferlito}, {Frenk}, \&
  et~al.}]{PakmorSpringelColes_2023MNRAS.524.2539P}
{Pakmor}, R., {Springel}, V., {Coles}, J.~P., {et~al.} 2023, \mnras, 524, 2539

\bibitem[{{Pandey} {et~al.}(2022){Pandey}, {Gatti}, {Baxter}, {Hill}, {Fang},
  {Doux}, {Giannini}, {Raveri}, {DeRose}, {Huang}, \&
  et~al.}]{PandeyGattiBaxter_2022PhRvD.105l3526P}
{Pandey}, S., {Gatti}, M., {Baxter}, E., {et~al.} 2022, \prd, 105, 123526

\bibitem[{{Peebles}(1974)}]{Peebles_1974ApJS...28...37P}
{Peebles}, P.~J.~E. 1974, \apjs, 28, 37

\bibitem[{{Pons-Border{\'\i}a} {et~al.}(1999){Pons-Border{\'\i}a},
  {Mart{\'\i}nez}, {Stoyan}, {Stoyan}, \&
  {Saar}}]{Pons-BorderiaMartinezStoyan_1999ApJ...523..480P}
{Pons-Border{\'\i}a}, M.-J., {Mart{\'\i}nez}, V.~J., {Stoyan}, D., {Stoyan},
  H., \& {Saar}, E. 1999, \apj, 523, 480

\bibitem[{{Popesso} {et~al.}(2024){Popesso}, {Marini}, {Dolag}, {Lamer},
  {Csizi}, {Biffi}, {Robothan}, {Bravo}, {Biviano}, {Vladutesku-Zopp}, \&
  et~al.}]{PopessoMariniDolag_2024arXiv241117120P}
{Popesso}, P., {Marini}, I., {Dolag}, K., {et~al.} 2024, arXiv e-prints,
  arXiv:2411.17120

\bibitem[{{Predehl} {et~al.}(2021){Predehl}, {Andritschke}, {Arefiev},
  {Babyshkin}, {Batanov}, {Becker}, {B{\"o}hringer}, {Bogomolov}, {Boller},
  {Borm}, \& et~al.}]{PredehlAndritschkeArefiev_2021A&A...647A...1P}
{Predehl}, P., {Andritschke}, R., {Arefiev}, V., {et~al.} 2021, \aap, 647, A1

\bibitem[{{Robertson} {et~al.}(2024){Robertson}, {Huff}, {Markovi{\v{c}}}, \&
  {Li}}]{RobertsonHuffMarkovic_2024MNRAS.533.4081R}
{Robertson}, A., {Huff}, E., {Markovi{\v{c}}}, K., \& {Li}, B. 2024, \mnras,
  533, 4081

\bibitem[{{Rykoff} {et~al.}(2014){Rykoff}, {Rozo}, {Busha}, {Cunha},
  {Finoguenov}, {Evrard}, {Hao}, {Koester}, {Leauthaud}, {Nord}, \&
  et~al.}]{RykoffRozoBusha_2014ApJ...785..104R}
{Rykoff}, E.~S., {Rozo}, E., {Busha}, M.~T., {et~al.} 2014, \apj, 785, 104

\bibitem[{{Salcedo} {et~al.}(2020){Salcedo}, {Wibking}, {Weinberg}, {Wu},
  {Ferrer}, {Eisenstein}, \&
  {Pinto}}]{SalcedoWibkingWeinberg_2020MNRAS.491.3061S}
{Salcedo}, A.~N., {Wibking}, B.~D., {Weinberg}, D.~H., {et~al.} 2020, \mnras,
  491, 3061

\bibitem[{{S{\'a}nchez} {et~al.}(2005){S{\'a}nchez}, {Lambas}, {B{\"o}hringer},
  \& {Schuecker}}]{SanchezLambasBohringer_2005MNRAS.362.1225S}
{S{\'a}nchez}, A.~G., {Lambas}, D.~G., {B{\"o}hringer}, H., \& {Schuecker}, P.
  2005, \mnras, 362, 1225

\bibitem[{{S{\'a}nchez} {et~al.}(2023){S{\'a}nchez}, {Omori}, {Chang}, {Bleem},
  {Crawford}, {Drlica-Wagner}, {Raghunathan}, {Zacharegkas}, {Abbott},
  {Aguena}, \& et~al.}]{SanchezOmoriChang_2023MNRAS.522.3163S}
{S{\'a}nchez}, J., {Omori}, Y., {Chang}, C., {et~al.} 2023, \mnras, 522, 3163

\bibitem[{{Schaye} {et~al.}(2023){Schaye}, {Kugel}, {Schaller}, {Helly},
  {Braspenning}, {Elbers}, {McCarthy}, {van Daalen}, {Vandenbroucke}, {Frenk},
  \& et~al.}]{SchayeKugelSchaller_2023MNRAS.526.4978S}
{Schaye}, J., {Kugel}, R., {Schaller}, M., {et~al.} 2023, \mnras, 526, 4978

\bibitem[{{Seppi} {et~al.}(2022){Seppi}, {Comparat}, {Bulbul}, {Nandra},
  {Merloni}, {Clerc}, {Liu}, {Ghirardini}, {Liu}, {Salvato}, \&
  et~al.}]{SeppiComparatBulbul_2022A&A...665A..78S}
{Seppi}, R., {Comparat}, J., {Bulbul}, E., {et~al.} 2022, \aap, 665, A78

\bibitem[{{Seppi} {et~al.}(2024){Seppi}, {Comparat}, {Ghirardini}, {Garrel},
  {Artis}, {S{\'a}nchez}, {Liu}, {Clerc}, {Bulbul}, {Grandis}, \&
  et~al.}]{SeppiComparatGhirardini_2024A&A...686A.196S}
{Seppi}, R., {Comparat}, J., {Ghirardini}, V., {et~al.} 2024, \aap, 686, A196

\bibitem[{{Sorini} {et~al.}(2024){Sorini}, {Bose}, {Dav{\'e}}, \&
  {Angl{\'e}s-Alc{\'a}zar}}]{SoriniBoseDave_2024OJAp....7E.115S}
{Sorini}, D., {Bose}, S., {Dav{\'e}}, R., \& {Angl{\'e}s-Alc{\'a}zar}, D. 2024,
  The Open Journal of Astrophysics, 7, 115

\bibitem[{{Springel} {et~al.}(2005){Springel}, {White}, {Jenkins}, {Frenk},
  {Yoshida}, {Gao}, {Navarro}, {Thacker}, {Croton}, {Helly}, {Peacock}, {Cole},
  {Thomas}, {Couchman}, {Evrard}, {Colberg}, \& {Pearce}}]{Springel2005}
{Springel}, V., {White}, S. D.~M., {Jenkins}, A., {et~al.} 2005, \nat, 435, 629

\bibitem[{{Sunyaev} {et~al.}(2021){Sunyaev}, {Arefiev}, {Babyshkin},
  {Bogomolov}, {Borisov}, {Buntov}, {Brunner}, {Burenin}, {Churazov},
  {Coutinho}, \& et~al.}]{SunyaevArefievBabyshkin_2021A&A...656A.132S}
{Sunyaev}, R., {Arefiev}, V., {Babyshkin}, V., {et~al.} 2021, \aap, 656, A132

\bibitem[{{Truong} {et~al.}(2021){Truong}, {Pillepich}, \&
  {Werner}}]{TruongPillepichWerner_2021MNRAS.501.2210T}
{Truong}, N., {Pillepich}, A., \& {Werner}, N. 2021, \mnras, 501, 2210

\bibitem[{{Truong} {et~al.}(2020){Truong}, {Pillepich}, {Werner}, {Nelson},
  {Lakhchaura}, {Weinberger}, {Springel}, {Vogelsberger}, \&
  {Hernquist}}]{TruongPillepichWerner_2020MNRAS.494..549T}
{Truong}, N., {Pillepich}, A., {Werner}, N., {et~al.} 2020, \mnras, 494, 549

\bibitem[{{van der Wel}(2008)}]{vanderWel_2008ApJ...675L..13V}
{van der Wel}, A. 2008, \apjl, 675, L13

\bibitem[{{Vani} {et~al.}(2025){Vani}, {Ayromlou}, {Kauffmann}, \&
  {Springel}}]{VaniAyromlouKauffmann_2025MNRAS.536..777V}
{Vani}, A., {Ayromlou}, M., {Kauffmann}, G., \& {Springel}, V. 2025, \mnras,
  536, 777

\bibitem[{{Vulcani} {et~al.}(2023){Vulcani}, {Poggianti}, {Gullieuszik},
  {Moretti}, {Fritz}, {Bettoni}, {Facciolli}, {Fasano}, \&
  {Omizzolo}}]{VulcaniPoggiantiGullieuszik_2023ApJ...949...73V}
{Vulcani}, B., {Poggianti}, B.~M., {Gullieuszik}, M., {et~al.} 2023, \apj, 949,
  73

\bibitem[{{Yan} {et~al.}(2021){Yan}, {van Waerbeke}, {Tr{\"o}ster}, {Wright},
  {Alonso}, {Asgari}, {Bilicki}, {Erben}, {Gu}, {Heymans}, \&
  et~al.}]{YanvanWaerbekeTroster_2021A&A...651A..76Y}
{Yan}, Z., {van Waerbeke}, L., {Tr{\"o}ster}, T., {et~al.} 2021, \aap, 651, A76

\bibitem[{{Yang} {et~al.}(2007){Yang}, {Mo}, {van den Bosch}, {Pasquali}, {Li},
  \& {Barden}}]{YangMovandenBosch_2007ApJ...671..153Y}
{Yang}, X., {Mo}, H.~J., {van den Bosch}, F.~C., {et~al.} 2007, \apj, 671, 153

\bibitem[{{York} {et~al.}(2000){York}, {Adelman}, {Anderson}, {Anderson},
  {Annis}, {Bahcall}, {Bakken}, {Barkhouser}, {Bastian}, {Berman}, \&
  et~al.}]{YorkAdelmanAnderson_2000AJ....120.1579Y}
{York}, D.~G., {Adelman}, J., {Anderson}, John~E., J., {et~al.} 2000, \aj, 120,
  1579

\bibitem[{{Zhang} {et~al.}(2024{\natexlab{a}}){Zhang}, {Comparat}, {Ponti},
  {Merloni}, {Nandra}, {Haberl}, {Locatelli}, {Zhang}, {Sanders}, {Zheng}, \&
  et~al.}]{ZhangComparatPonti_2024A&A...690A.267Z}
{Zhang}, Y., {Comparat}, J., {Ponti}, G., {et~al.} 2024{\natexlab{a}}, \aap,
  690, A267

\bibitem[{{Zhang} {et~al.}(2024{\natexlab{b}}){Zhang}, {Comparat}, {Ponti},
  {Merloni}, {Nandra}, {Haberl}, {Truong}, {Pillepich}, {Locatelli}, {Zhang},
  \& et~al.}]{ZhangComparatPonti_2024A&A...690A.268Z}
{Zhang}, Y., {Comparat}, J., {Ponti}, G., {et~al.} 2024{\natexlab{b}}, \aap,
  690, A268

\bibitem[{{Zhang} {et~al.}(2025){Zhang}, {Comparat}, {Ponti}, {Merloni},
  {Nandra}, {Haberl}, {Truong}, {Pillepich}, {Popesso}, {Locatelli}, \&
  et~al.}]{ZhangComparatPonti_2025A&A...693A.197Z}
{Zhang}, Y., {Comparat}, J., {Ponti}, G., {et~al.} 2025, \aap, 693, A197

\bibitem[{{Zu} \& {Mandelbaum}(2015)}]{ZuMandelbaum_2015MNRAS.454.1161Z}
{Zu}, Y. \& {Mandelbaum}, R. 2015, \mnras, 454, 1161

\bibitem[{{Zu} \& {Mandelbaum}(2016)}]{ZuMandelbaum_2016MNRAS.457.4360Z}
{Zu}, Y. \& {Mandelbaum}, R. 2016, \mnras, 457, 4360

\bibitem[{{Zu} \& {Weinberg}(2013)}]{ZuWeinberg_2013MNRAS.431.3319Z}
{Zu}, Y. \& {Weinberg}, D.~H. 2013, \mnras, 431, 3319

\end{thebibliography}

\end{document}